# Current-induced magnetization control in dipolar-coupled nanomagnet pairs and artificial spin ice

A. Pac[1,2*], G. M. Macauley[1,2,†], J. A. Brock[1,2], A. Hrabec[1,2], A. Kurenkov[1,2], V. Raposo[3,4], E. Martinez[3,4], and L. J. Heyderman[1,2*]

1. Laboratory for Mesoscopic Systems, Department of Materials, ETH Zurich, 8093 Zurich, Switzerland
2. PSI Center for Neutron and Muon Sciences, 5232 Villigen PSI, Switzerland
3. Department of Applied Physics, University of Salamanca, 37008 Salamanca, Spain
4. Unidad de Excelencia en Luz y Materia Estructuradas (LUMES), Universidad de Salamanca, Salamanca, Spain

Corresponding authors: aleksandra.pac@psi.ch, laura.heyderman@psi.ch

† Present Address: Department of Physics, Princeton University, Princeton, NJ 08540 USA

## Abstract

Exploiting current-induced spin-orbit torques (SOTs) to manipulate the magnetic state of dipolar-coupled nanomagnet systems with in-plane magnetic anisotropy, such as artificial spin ices, provides a route to local, electrically-programmable control of the magnetization, with relevance for applications including neuromorphic computing. Here, we demonstrate how the orientation of a nanomagnet relative to the direction of an applied electrical current impacts the threshold current density needed for all-electrical magnetization switching, and how dipolar coupling between the nanomagnets influences the switching of interacting pairs and ensembles of nanomagnets. Using a material system designed to generate SOTs in response to electrical currents, we find that the current required to switch the magnetization of isolated nanomagnets varies non-monotonically as the angle between the nanomagnet long axis and the current increases. In small artificial spin ice systems, we observe similar angular dependence of the switching current, which can be used to control the magnetization orientation of specific subsets of nanomagnets. These experimental results are supported by micromagnetic modeling, which illustrates how the various current-induced torques can be exploited to control magnetization switching in nanomagnetic systems. These results establish SOT switching as a practical method for programmable manipulation of dipolar nanomagnetic systems.

I.  **Introduction**

Artificial spin ices, which are arrays of coupled mesoscopic, single-domain magnets, were first introduced as two-dimensional (2D) analogues of frustrated magnetic systems [1,2]. Since then, they have proven to be a powerful platform for exploring fundamental questions involving phase transitions [3–7] and emergent magnetic monopoles [8,9]. Increasingly, they are also being exploited in schemes for low-power computation, such as reservoir and neuromorphic computing [10–16], as well as in reconfigurable magnetic logic [17–19].

Controlling the precise microstate of artificial spin ices, however, remains an outstanding challenge. Applied magnetic fields have provided the most direct means of reconfiguring their magnetic state [1,20,21], with the field orientation enabling specific magnetization reversal processes, including those mediated by emergent domain walls and clocked dynamics [20,21]. In addition, thermal activation has been used to promote relaxation to low-energy states across a variety of lattices [6,22–24]. More recently, plasmon-assisted photo-heating has offered spatial selectivity beyond global heating, enabling targeted reversal of individual nanomagnets within nanomagnetic logic gates [25]. Setting the state of individual nanomagnets can also be achieved using a scanning magnetic tip, though this approach is inherently slow [26]. For scalable and energy-efficient applications, it is vital to have local control and read-out of the magnetic state through purely electrical means, i.e. by applying an electric current [27].

Current-driven control of in-plane nanomagnets—both as individual elements and within dipolar-coupled arrays of nanomagnets—opens entirely new vistas for controlling magnetization reversal and for realizing neuromorphic computing architectures. Here, spin-orbit torques (SOTs) have emerged as a powerful mechanism to electrically control the magnetization within magnetic thin films and tunnel junctions [28–31]. SOTs arise when an electrical current passes through a heavy-metal/ferromagnet heterostructure. The electrical current in the heavy metal gives rise to a transverse spin current (of spin polarization $\sigma$) that can exert a torque on the magnetic moment $m$ in the adjacent ferromagnetic layer, and the resulting dynamics depends sensitively on the orientation between the electric current and the local magnetization [32]. Deterministic switching requires an additional symmetry-breaking mechanism when $\sigma$ is orthogonal to $m$, typically achieved by applying a magnetic field orthogonal to both $\sigma$ and $m$ [31]. This constraint has hindered the widespread deployment of SOTs in devices based on out-of-plane magnetized films although this mechanism has been demonstrated in a few individual cases [33]. In contrast, in-plane magnetized systems [27,31] inherently permit the alignment between $m$ and $\sigma$ [27], enabling fully deterministic switching in certain scenarios.

In this work, we demonstrate current-induced switching of isolated nanomagnets and coupled nanomagnets that are in pairs and arrays of nanomagnets arranged on a square lattice referred to as artificial square ice. We establish how their orientation with respect to the electric current impacts the threshold currents required for all-electrical magnetization reversal. By optimizing the magnitude and direction of the applied current, we demonstrate control of the magnetic state of small artificial spin ice systems via specific magnetization reversal paths.

We begin by quantifying the current-induced switching of isolated nanomagnets, extracting the minimum threshold current density required to switch the magnetization as a function of the angle between the nanomagnet long axis and the current direction. We then examine how dipolar interactions modify the switching behavior of pairs of nanomagnets. Finally, we extend this analysis to artificial square ice lattices composed of orthogonal sublattices of nanomagnets.

Together, these results establish a framework for current-driven control of artificial spin ices and provide a basis for their use in spintronics, for example, as part of reconfigurable magnonics and neuromorphic architectures.

II. **Current-Induced Magnetization Switching of Isolated Nanomagnets**

We first examine the SOT switching of isolated nanomagnets as a basis for understanding the collective behavior that emerges in coupled nanostructures. In our experiments, individual stadium-shaped permalloy (Py) nanomagnets are placed on top of a Pt channel, which is schematically shown in Fig. 1(a). Each nanomagnet has a length *l* = 450 nm, width *w* = 150 nm, and thickness *t* = 10 nm. The large in-plane aspect ratio, $l/w > 1$, imparts a strong shape anisotropy which, combined with their small dimensions, ensures that each nanomagnet remains single-domain, with its magnetization $\boldsymbol{m}$ aligned along its long axis. In general, we can therefore neglect the internal magnetic texture within individual nanomagnets, and describe their behavior in terms of a single macrospin, representing the uniform magnetization within each element.

The time evolution of the magnetization $\boldsymbol{m}$ under the influence of various torques is described by the Landau-Lifshitz-Gilbert (LLG) equation,

$$\frac{\partial \boldsymbol{m}}{\partial t} = -\gamma \boldsymbol{m} \times \boldsymbol{H}_{\text{eff}} + \alpha \boldsymbol{m} \times \frac{\partial \boldsymbol{m}}{\partial t} + \boldsymbol{\tau}_{\text{SOT}}. \tag{1}$$

where $\gamma$ is the gyromagnetic ratio and $\alpha$ is the dimensionless Gilbert damping constant. The first term on the right-hand side of Eq. (1), leads to precessional motion of $\boldsymbol{m}$ around the direction of effective magnetic field $\boldsymbol{H}_{\text{eff}}$. The second, phenomenological term causes the magnetization to eventually align with the direction of the effective magnetic field. The third term represents the contribution of spin-orbit torques $\boldsymbol{\tau}_{\text{SOT}}$ to the behaviour of $\boldsymbol{m}$.

When an electric current, with current density $J_\text{C}$ passes through the Pt layer, the spin Hall effect generates a transverse spin current (of spin polarization $\boldsymbol{\sigma}$), as indicated in Fig. 1(a) [34–36]. This results in a spin accumulation at the Pt/Py interface that exerts torques on the magnetization within the nanomagnets. In general, the resulting spin-orbit torques can be decomposed into two perpendicular components: a field-like torque $\boldsymbol{\tau}_{\text{FL}}$, which causes precessional motion of the magnetization, and a damping-like torque $\boldsymbol{\tau}_{\text{DL}}$, which promotes reversal of $\boldsymbol{m}$. These torques can be expressed as

$$\boldsymbol{\tau}_{\text{FL}} = \tau_{\text{FL}} \boldsymbol{\sigma} \times \boldsymbol{m}, \text{ and} \tag{2.1}$$

$$\boldsymbol{\tau}_{\text{DL}} = \tau_{\text{DL}} (\boldsymbol{m} \times (\boldsymbol{m} \times \boldsymbol{\sigma})), \tag{2.2}$$

where $\tau_{\text{FL}}$ and $\tau_{\text{DL}}$ are their associated magnitudes [32,37,38]. Equivalently, the torques can be expressed in terms of the cross product of $\boldsymbol{m}$ with effective field-like and damping-like magnetic fields, defined as $\mu_0 \boldsymbol{H}_{\text{FL}} = -\tau_{\text{FL}} \boldsymbol{\sigma}$ and $\mu_0 \boldsymbol{H}_{\text{DL}} = \tau_{\text{DL}} (\boldsymbol{m} \times \boldsymbol{\sigma})$, respectively. In addition to the effective fields resulting from spin-orbit torques, the applied current gives rise to an Oersted field, which acts in the opposite direction to the field-like effective field for a Pt layer [39]. The orientations of the effective fields $\mu_0 \boldsymbol{H}_{\text{FL}}$ and $\mu_0 \boldsymbol{H}_{\text{DL}}$, and the Oersted field $\mu_0 \boldsymbol{H}_{\text{Oe}}$, are shown in Fig. 1(a). Typically, the strengths of the associated torques depend on the intrinsic properties of the heavy metal and magnetic material, their respective thicknesses, and the quality of the interface between them [40]. The strength of the damping-like torque is usually expressed in terms of the spin Hall angle $\theta_{SH}$, for which we assume the value of +0.2 for Pt, according to

$$\tau_{DL} = \frac{\theta_{SH}\hbar}{2eM_S t_{FM}} J_C, \quad (3)$$

where $M_S$ is the saturation magnetization of the magnetic layer, $t_{FM}$ is its thickness, $e$ is the electron charge and $\hbar$ is Planck's constant.

It is conventional to distinguish between two geometries for SOT-induced switching of in-plane magnetized nanomagnets, defined by the orientation of the initial magnetization relative to the direction of the current density. In both of these geometries, the magnetization of the ferromagnetic component—here the macrospin associated with the single-domain net moment of the nanomagnet—lies in-plane. Based on the coordinate system in Fig. 1(a), in which the current density $J_C$ flows along x-axis, these two types are:

- Type *x*, in which the long axis of the nanomagnet, and therefore the initial magnetization, is collinear with, and either parallel or antiparallel to, the applied current [27]; and
- Type *y*, in which the long axis of the nanomagnet, and therefore the initial magnetization, is orthogonal to the applied current [31].

In Fig. 1(a), the initial magnetic configuration of the nanomagnet is represented by the macrospin $m$, which is oriented at an angle $\phi \leq 90°$ with respect to the negative x-axis. In what follows, we correspondingly define $\phi \in [0°, 90°]$ as the angle between the long axis of the nanomagnet and the channel. In each figure, the current direction will be indicated explicitly to uniquely specify the magnetic configuration with respect to the current flow.

We patterned isolated Py nanomagnets with various orientations $\phi$ on top of Pt channels using electron-beam lithography in combination with ion-beam milling of a Ta(3 nm)/Pt(6 nm)/Py(10 nm)/Ru (2 nm) sputter-deposited multilayer stack. The thickness of the Py layer was chosen to be large enough to provide sufficient magnetic contrast when imaging the nanomagnets using magnetic force microscopy (MFM). The Pt channels are connected to Cr(5 nm)/Au(100 nm) electrodes patterned on top of the sputtered magnetic stack and deposited by thermal evaporation. Scanning electron microscopy (SEM) images of the entire device layout are provided in the Supplemental Materials, Section 1.

To determine the impact of the orientation of the applied current on the threshold current $J_{switch}$ required to switch the magnetization we fabricated nanomagnets with $\phi$ increasing from 0° to 90° in 15° increments. For each orientation, five nanomagnets were patterned on a channel. Each nanomagnet on this channel is separated from its immediate neighbors by 2.5 μm to reduce the dipolar coupling between the nanomagnets. An SEM image of a channel with nanomagnets oriented at $\phi = 90°$ to the current direction is shown in Fig. 1(b). We label the five nanomagnets from left to right ① to ⑤. Also visible are horizontal slits, which separate adjacent channels where nanomagnets of different orientations are patterned. A detailed rationale of the chosen sample design is provided in the Supplemental Materials, Section 1.

The nanomagnets were first initialized by applying an in-plane magnetic field to set them all in in the same magnetic configuration. This was confirmed using MFM [first row of Fig. 1(c) for nanomagnets with $\phi = 90°$], where the identical orientation of bright-dark contrast for every nanomagnet confirms that all the macrospins point in the same direction as indicated by the green arrows.

After initializing the magnetic state, 1-µs-long current pulses of increasing amplitude were applied using a pulse generator, with no reinitialization of the magnetic state between pulses. The resulting magnetic configuration after each pulse was imaged using MFM. The current density was gradually increased from $4.86 \times 10^{12}$ Am$^{-2}$ to $5.38 \times 10^{12}$ Am$^{-2}$, in steps of $\sim 0.05 \times 10^{12}$ Am$^{-2}$. For the $\phi = 90°$ nanomagnets shown in Fig. 1(c), nanomagnet ③ was the first to switch, followed by nanomagnet ②, and then nanomagnet ④. No switching was observed for either nanomagnet ① or nanomagnet ⑤, even when the current is further increased, because of reduced current density due to fringing at both sides of the channel (Supplemental Materials, Section 1).

We show the dependence of the threshold current density for switching $J_{switch}$ on $\phi$ in Fig. 1(d). Each data point is the average of the minimum current density required to switch the three central nanomagnets, i.e., ②, ③ and ④. We observe a maximum in $J_{switch}$ near $\phi = 0°$, where the switching is primarily of Type *x*, since the long axis of the nanomagnet and the current density are parallel. On increasing $\phi$, the threshold current density first decreases until it reaches a minimum at $\phi = 75°$ before it increases again at $\phi = 90°$, by which point the switching is of Type *y*, because the macrospin and current direction are orthogonal. The increase in $J_{switch}$ for $\phi = 90°$ is somewhat different to the previous experimental observation of a monotonic dependence of $J_{switch}$ on $\phi$ [27].

To aid in explaining the origin of the angular dependence observed here and to assess the role of the Oersted field, we performed COMSOL simulations to estimate the magnitude of the Oersted field (see Supplemental Materials, Section 2). We further investigated the switching fields and energy barriers of both isolated and pairs of nanomagnets using micromagnetic simulations (see Supplemental Materials, Section 3). From the simulations, we observe a minimum in the switching field at an angle of ~22°. Despite the fact that the simulations for magnetic field driven switching, this is in reasonable agreement with the experimentally obtain value of ~15° obtained from current-induced switching shown in Fig. 1(d). These results indicate that the location of the minimum in $J_{switch}$ ($\phi$) is governed by multiple factors, including the nanomagnet aspect ratio, the magnitude of the Oersted field, and the relative strengths of the field-like and damping-like effective fields (see Supplemental Materials, Section 6).

Assigning a precise value for the threshold current density is difficult in the case of pure Type *x* switching. In this geometry, with $\phi = 0°$, the long axis of the nanomagnet is aligned with the current density. For sufficiently high current densities, the field-like component of the SOT drives the magnetization to point along the short axis of the nanomagnet. When the current is removed, and in the absence of any intrinsic disorder that may break its symmetry, the macrospin has an equal probability of relaxing into either of the two minimum energy states corresponding to the macrospin pointing in one of two directions parallel to the long axis of the nanomagnet [41]. This may mean that the macrospin may relax back to its initial state, even though sufficient impetus was provided to switch. To differentiate this type of behavior from deterministic switching observed at other angles, the point at $\phi = 0°$ is represented by an unfilled marker. It corresponds to the current density at which the magnetization first switches to an opposite direction from its initial magnetization direction after a current pulse, although this measurement may not reflect the lowest switching current density $J_{switch}$. The current-induced reversal has been verified with two further experimental geometries for switching the magnetization of isolated nanomagnets; (i) reversing the current direction or (ii) reversing both the initial magnetization direction and the current direction (see Supplemental Materials, Section 5).

To complement our experimental data, we analyzed the dependence of $J_{switch}$ on $\phi$ using micromagnetic simulations [Fig. 1(e)]. In the simulations, the field-like torque is assumed to be proportional to the damping-like torque, expressed as $H_{FL} = kH_{DL}$. The exact parameters used for the simulations can be found in the Supplemental Materials, Section 6. We compare the experimental switching trend from Fig. 1(d) with the simulated results in Fig. 1(e), which include the Oersted field and a ratio of the field-like component to the damping-like component of $k = 0.5$. For $k = 0.5$, the simulations display a minimum in $J_{switch}$ around 90°, while the experimental data show a minimum in $J_{switch}$ at $\phi \sim 75°$; for the experimental system, $J_{switch}$ begins to increase while, for the simulated system, $J_{switch}$ remains constant before decreasing further at 90°. We performed further simulations with and without the Oersted field but now with $k = 0$, which represents the inclusion of a purely damping-like torque – commonly considered the primary driver of magnetization switching by stabilizing the reversed state and minimizing total energy [27,42]. – The simulated minimum shifts to 85°–90° when Oersted field is included, while the minimum shifts to larger values when Oersted field is not considered. These simulations are summarized in Supplemental Materials, Section 6.1. These results indicate that in our system, all contributions—the field-like torque, damping-like torque, and the Oersted field—determine the switching behavior observed in Fig. 1(d).

Slightly larger nanomagnets ($l$ = 470 nm, $w$ = 170 nm) can support either a single-domain state or a vortex state [43,44]. Interestingly, for such nanomagnets at $\phi = 0°$, the circularity of the vortex state [10]—clockwise or anticlockwise—can be deterministically controlled, providing a means to toggle between them, despite the fact that the initial state was single domain after the initialization with the magnetic field. This is illustrated in the upper row of images in Fig. 1(f). The threshold current density required for toggling the vortex chirality $J_T$ is $6.08 \times 10^{12}$ Am$^{-2}$, exceeding the current density required to switch the smaller nanomagnets at $\phi = 0°$, which is around $5.1 \times 10^{12}$ Am$^{-2}$. However, for nanomagnets oriented at $\phi = 15°$, the switching of the vortex chirality becomes less deterministic, with current pulses leading to a mixture of vortex, multidomain, and single-domain states [see lower row of images in Fig. 1(f)].

### III. Switching of pairs of nanomagnets on current injection

Now that we have established the current-induced switching behavior of isolated nanomagnets, we turn to examining the simplest example of a coupled system, namely, a pair of dipolar-coupled nanomagnets. We fabricated devices with two different arrangements of pairs: either (i) side-by-side [shown in Fig. 2(a)] or (ii) end-to-end [shown in Fig. S1.1(c), Supplemental Materials]. The edge-to-edge separation of the two nanomagnets arranged side-by-side (end-to-end) is 18 nm (22 nm). In both cases, this was the minimum spacing—and hence largest dipolar coupling—we could achieve with our fabrication process.

To gain an insight into the effect of dipolar interaction on the switching characteristics, we first performed micromagnetic simulations to estimate the intrinsic energy barrier to magnetization reversal. For simplicity, we adopted the Stoner-Wohlfarth model [45], in which the macrospin associated with a nanomagnet rotates in a coherent fashion between two bistable configurations separated by an energy barrier. The macrospin rotation is parameterized by an angle $\theta$ (relative to the long axis). In this picture, the energy minima correspond to the situation in which the macrospin is aligned along the long axis of the nanomagnet, while the energy maximum—corresponding to the energy barrier with respect to the energy of the initial configuration—occurs when the macrospin points along the short axis.

We show schematics for the possible switching processes for the side-by-side pair in Fig. 2(b), with the rotation angles for the macrospins in two nanomagnets, $\theta_1$ and $\theta_2$, indicated. For two side-by-side nanomagnets, the lowest-energy state is the one in which the macrospins are antiparallel. When both macrospins are aligned parallel—as occurs when an external field is used to initialize them—the state is energetically metastable. We identify several pathways for reversing the macrospins from one parallel state to the other:

1. *Synchronous switching,* in which both nanomagnets reverse in a single step. In a simplified model, we assume that both nanomagnets reverse at the same rate, and they can either reverse both in the same sense [blue panel, Fig. 2(b)] or in opposite senses [orange panel, Fig. 2(b)].
2. *Sequential switching*, in which one nanomagnet reverses [green panel, Fig. 2(b)], with the second nanomagnet reversing at a later time.

In Fig. 2(c), we show the energy as a function of the macrospin angle for different scenarios. For a single, isolated nanomagnet, the energy barrier to switching is found to be ~179 eV [black curve, Fig. 2(c)]. In the case of *sequential* switching shown in the green panel of Fig. 2(b), the reversal of the first nanomagnet happens via clockwise or anticlockwise rotation of the magnetization while the magnetization in the second nanomagnet remains unchanged and this corresponds to the green curve in Fig. 2(c) with an energy barrier of 147 eV. The reversal of the second nanomagnet then follows the same green curve shown in Fig. 2(c), but this time in reverse, so that the energy barrier is increased to 182.5 eV with the energy associated with the final parallel state of the system higher than that of the antiparallel configuration.

We now consider the situation in which both nanomagnets switch in a single step, which we refer to as *synchronous* switching. For simplicity, we assume that both macrospins rotate at the same rate, so that $\theta = |\theta_1| = |\theta_2|$, and we find that the energy barrier is significantly higher than for sequential switching. When the macrospins rotate in the same direction (both clockwise or both anticlockwise), the energy barrier is 261 eV [blue panel and curve in Fig. 2(b) and 2(c), respectively]. When they rotate in opposite directions, this energy barrier increases to 385 eV [orange panel and curve in Fig. 2(b) and 2(c), respectively]. In both of these scenarios for synchronous switching, the initial and final states have the same energy, as evidenced by the symmetry of the blue and orange curves. Due to the much larger energy barrier for the synchronous switching, we would expect to observe sequential switching experimentally.

For an end-to-end pair of nanomagnets, the initial parallel configuration after applying a magnetic field is the system's low-energy state, and the antiparallel configuration is energetically unfavorable, making synchronous switching of the end-to-end nanomagnet pair more likely. The simulation results for the end-to-end pair of nanomagnets are discussed in detail in the Supplemental Materials, Section 3.

Now that we have established the expected behavior for nanomagnetic pairs, we turn to the experimental results. The procedure used was the same as for the isolated nanomagnets. We begin by initializing the nanomagnets into a parallel configuration using a magnetic field and then applied current pulses of increasing amplitude. As predicted, for the side-by-side nanomagnet pair, the reversal process is sequential and, in Fig. 2(d), we show the experimental results for $J_{switch}$ versus $\phi$ for the first and second step, which are shown in light and dark purple, respectively. We see in Fig. 2(d) that the behavior for the side-by-side pairs of nanomagnets is qualitatively the same as that for isolated nanomagnets. In particular, $J_{switch}$ has a minimum at around $\phi = 75°$, with a gradual increase in $J_{switch}$ with the increase or decrease of the angle. This trend in $J_{switch}$ versus $\phi$ is also seen for end-to-end pairs of nanomagnets (see Supplemental Materials, Section 4).

Assuming that the two nanomagnets are identical, for the first step, going from a parallel to antiparallel configuration, there should be no preference for which nanomagnet switches first and, therefore, $J_{switch}$ corresponds to the current required to switch either nanomagnet in the pair. Nevertheless, any difference between the nanomagnets that affects in the energy barriers to switching, could lead to a more deterministic switching sequence within the pair. In accordance with the micromagnetic simulations, where the energy barrier associated with switching the first nanomagnet is lower than the energy barrier to switching of an isolated nanomagnet, it can be seen in Fig. 2(d) that, in general, the threshold current densities for the first step are lower than those required to switch the isolated nanomagnets, and are also lower than those required for the second step. This is particularly evident at higher angles.

We verify our experimental observations with micromagnetic simulations of SOT switching for side-by-side pairs of magnets for different $J_{switch}$ and $\phi$ as shown in Fig. 2(e). The "figure of merit" reflects the magnetic configuration and represents the dot product of the initial and final magnetization states for each nanomagnet in the pair summed together (**m1**$_{start}$·**m1**$_{end}$ + **m2**$_{start}$·**m2**$_{end}$), where **m** = +1 and -1 corresponds to the nanomagnet magnetization before and after switching. At current densities insufficient to switch the nanomagnets, the figure of merit is +2 and corresponds to the yellow region in Fig. 2(e). At sufficiently high current densities, both nanomagnets switch, and the figure of merit is -2. Between these two regions lies a narrow green region, corresponding to the situation in which only one of the nanomagnets has switched. It can be seen in Fig. 2(e) that, for most angles, as $J_{switch}$ increases, the system transitions from no switching (yellow) to one nanomagnet switching (green), and then to both nanomagnets switching (dark blue). This is consistent with the experimental results and our estimates of the energy barriers; namely that for nearly all $\phi$, on increasing $J_{switch}$, the green region in Fig. 2(e) is reached before the dark blue region, and magnetization reversal proceeds by sequential switching of the nanomagnets. This indicates that, with judicious choice of current amplitude, one can control the final magnetic configuration of a pair of dipolar-coupled nanomagnets.

From the angular dependence of $J_{switch}$ for end-to-end pairs of nanomagnets, we find that sequential switching is rarely observed experimentally and that simultaneous switching of both nanomagnets via a single step is more common. As mentioned above, this is because, starting from the initial parallel configuration, the system is more likely to switch directly to the opposite parallel configuration, than to the energetically unfavorable antiparallel configuration. Both the simulation and experimental results for the end-to-end pairs are discussed in the Supplemental Materials, Sections 3 and 4, respectively.

IV. **Switching of artificial spin ice structures**

Having demonstrated deterministic switching in pairs of dipolar-coupled nanomagnets, we now extend our approach to artificial square ices, which consist of small arrays of stadium-shaped nanomagnets arranged on the square lattice [1] [Fig. 3(a)]. Its magnetic behavior, under applied magnetic fields [1,46,47], or when thermally active [22,48], is among the most thoroughly explored of artificial spin ices. With its two orthogonal sublattices of nanomagnets and well-defined hierarchy of interactions, the artificial square ice is an ideal testbed for probing how SOTs can be exploited to control the magnetic configuration. A representative

SEM image of a typical artificial square ice from one of our devices is shown in Fig. 3(a). Using the same channel design employed for the single-nanomagnet measurements, the largest artificial square ice that can be accommodated between the slits in each region comprises a 4 × 4 array of nanomagnets.

In Fig. 3(b), we illustrate in a simplified schematic the mechanism for the switching of two of the nanomagnets in one of the square rings with a current applied in the horizontal direction as indicated. The initial magnetic state is set with an applied magnetic field $\mu_0 H_{ext}$, and we indicate the macrospin direction with green arrows. The vertical nanomagnets will switch for a sufficiently large current density due to a combination of $\mu_0 H_{FL}$, $\mu_0 H_{DL}$ and the Oersted field, with the effective vertical field indicated with purple arrows, while the current density it is not large enough to switch the magnetization in the horizontal magnets. However, it should be mentioned that, when the current density is sufficiently large so that the magnetization in the horizontal magnets points long the short axis, the nanomagnets can reverse with a certain probability.

We know from the experimental results shown in Fig. 1(d) that $J_{switch}$ for nanomagnets with $\phi = 90°$ is lower than $J_{switch}$ for nanomagnets near to $\phi = 0°$. Therefore, nanomagnets with their long axis orthogonal to the current are expected to switch at a lower current density than the nanomagnets with their long axis parallel to the current. As expected, our experimental measurements show that the $\phi = 90°$ nanomagnets switch first when we apply negative-polarity current [see Fig. 3(c)]. In the initialized state, magnetization of the Type *y* (Type *x*) nanomagnets points down (to the right), as indicated by the green arrows overlaid on the first MFM image in Fig. 3(c). A current pulse of negative polarity is applied, as indicated above the MFM images as well as the current pulse amplitude. Six current pulses with a width of 1 μs were applied, increasing in amplitude from $4.95 \times 10^{12}$ Am$^{12}$ to $5.38 \times 10^{12}$ Am$^{-2}$, while the magnetic state was measured after each pulse using MFM, without re-initializing the magnetic state in between the current pulses. The magnetization of the nanomagnets after switching is shown with purple arrows.

For the experiment shown in Fig. 3(c), since a negative current is applied along the horizontal direction, this leads to the reversal of nanomagnets in the vertical sublattice, beginning at the rightmost edge of the array. The Type *y* geometry of the vertical nanomagnets, with their long axis orthogonal to the current direction, means that switching is deterministic, without the need for an applied magnetic field. As the current density is increased, vertical nanomagnets within the artificial spin ice progressively switch. It is not clear why there is a progressive switching of the vertical nanomagnets from right to left, which may be a result of local variations in geometry [17], dipolar interactions [49] or current density.

The horizontal nanomagnets, which have a Type *x* geometry, with their long axis (approximately) parallel to the direction of the applied current, will also switch when the magnitude of the current density is sufficient. Indeed, we find that, at a current density greater than $5.21 \times 10^{12}$ Am$^{-2}$, nanomagnets in the horizontal sublattice begin to switch. As seen for the vertical nanomagnets, the number of nanomagnets that have switched in the horizontal sublattice increases with increasing current amplitude. The progression in switched nanomagnets with increasing current density is displayed quantitatively in Fig. 3(e) where the cumulative number of switches for Type *x* and Type *y* magnets is shown. Here, a distinct gap in the threshold $J_{switch}$ at which switching of horizontal and vertical nanomagnets starts is evident. Interestingly, Type *x* nanomagnets start switching at lower current densities ($5.21 \times 10^{12}$ Am$^{-2}$) than those required to switch isolated nanomagnets ($5.28 \times 10^{12}$ Am$^{-2}$) and side-by-side pairs ($5.35 \times 10^{12}$ Am$^{-2}$). Again, this reduction in the switching current may be related to the modified local magnetic field environment. From these experiments, we can infer that, for a large enough current pulse, both sublattices will switch simultaneously after a single current pulse.

When a current with positive polarity is applied, no switching is observed in the vertical nanomagnets [see Fig. 3(d)], because the direction of the effective field is aligned with the initial magnetization. As expected, switching of the horizontal nanomagnets is not observed until higher current densities are reached (+5.38 × $10^{12}$ A m$^{-2}$). Even then, the number of switched horizontal nanomagnets for positive current polarity is significantly smaller than that observed for negative polarity. Since the current-induced effective field in the horizontal nanomagnets is directed along their short axis, one would expect the onset of switching to occur at comparable current densities for both polarities. This observed asymmetry may be a result of slight differences in the pulse shapes associated with the positive and negative current pulses.

The experimental results obtained for the artificial square ice demonstrate that the switching behavior can be effectively defined by the nanomagnet orientation—Type *x* or Type *y*—and the polarity of the applied current. In terms of the geometry, for negative current, a notable difference is observed in the onset of the first switching events between Type *y* ($J$ = -4.95×$10^{12}$ Am$^{-2}$) and Type *x* ($J$ = -5.21×$10^{12}$ Am$^{-2}$). This indicates that sublattice control can be achieved by applying a global input in the form of electric currents. Such control introduces new possibilities for manipulating magnetization reversal in artificial spin systems composed of stadium-shaped nanomagnets.

Finally, we investigate whether specific magnetic states in small arrays can be achieved with a single current pulse or if a gradual increase in pulse amplitude is required to progressively modify the magnetic configuration. To address this, we compare (i) the magnetic configurations obtained on *continuously* increasing the current density [Fig. 6(a), top panel], based on the experiment shown in Fig. 3(a), with (ii) the magnetic configurations achieved on injecting a *single* pulse of the same amplitude, but this time initializing the system with an applied magnetic field between each current pulse [Fig. 4(a), bottom panel]. We then determine the percentage overlap, which is the percentage of nanomagnets that have the same orientation of the magnetization at a given current for the two different procedures. This is shown for Type *y* and Type *x* nanomagnets in Fig. 4(b). Across all current densities, the overlap remains above 80% for both nanomagnet types, indicating that the overall magnetic configurations achieved on applying a single pulse and a series of pulses with increasing current density are very similar. Therefore, a single current pulse is sufficient to modify the magnetic state into a specific configuration, independent of the initial magnetic state.

V.  **Conclusion**

We have demonstrated deterministic, field-free magnetization reversal in single-domain Py nanomagnets driven by SOTs from an underlying Pt layer. By systematically comparing isolated nanomagnets, dipolar-coupled pairs, and small artificial square ices consisting of arrays of dipolar-coupled nanomagnets, we have established how geometry and collective interactions govern the threshold current density required for switching.

For isolated nanomagnets (450 nm × 150 nm × 10 nm), the minimum threshold current density occurred near ~75°, with successful switching observed for angles between 15° and 90° at current densities in the range 4.9 - 5.4 × $10^{12}$ Am$^{-2}$. For nanomagnet pairs in the side-by-side configuration, switching occurred in two steps: starting from metastable parallel alignment, one nanomagnet reversed first giving an antiparallel alignment and stabilizing the system in a global energy minimum. This was followed by a second step at a slightly higher current density as the pair of nanomagnets transitioned from an antiparallel configuration back to the opposite metastable parallel configuration. In contrast, for end-to-end pairs of nanomagnets, magnetization reversal occurred in a single switching step with both nanomagnets reversing synchronously rather than sequentially.

Extending this concept to arrays of the artificial square ice, we showed that the two orthogonal nanomagnet sublattices—one with magnetization parallel to the current, and the other with magnetization perpendicular to the current—exhibit distinct reversal thresholds. This enables selective, current-driven control of one nanomagnet sublattice while leaving the orthogonal sublattice unaffected.

Together, these results introduce SOTs as a practical route for controlling artificial spin ices through purely electrical means, offering a straightforward approach that is compatible with existing CMOS architectures. This is an important step towards using artificial spin ices arrays as platforms for programmable magnetic logic and neuromorphic architectures.

**Acknowledgements**

We thank Anja Weber for her help with sample fabrication, Thomas Jung for access to the MFM tool, and Simone Finizio for his help with preparing the PCB for electrical setup. We are grateful for financial support from the Swiss National Science Foundation (project no. 200020_200332). We would also like to acknowledge the support from the EU FET-Open RIA project SpinENGINE (Grant no. 861618). JAB acknowledges support from the European Union's Horizon 2020 research and innovation programme under the Marie Skłodowska-Curie grant agreement No 884104 (PSI-FELLOW-III-3i). The work by V.R. and E.M. was partially supported by Project No. PID2023-150853NB-C31 funded by MICIU/AEI/10.13039/501100011033 and by FEDER, UE. Financial support of the Department of Education of the Junta de Castilla y León and FEDER Funds is gratefully acknowledged (Escalera de Excelencia CLU-2023-1-02).

**References**

[1] R. F. Wang et al., Artificial 'spin ice' in a geometrically frustrated lattice of nanoscale ferromagnetic islands, Nature **439**, 303 (2006).

[2] S. H. Skjærvø, C. H. Marrows, R. L. Stamps, and L. J. Heyderman, Advances in artificial spin ice, Nat. Rev. Phys. **2**, 13 (2020).

[3] L. Anghinolfi, H. Luetkens, J. Perron, M. G. Flokstra, O. Sendetskyi, A. Suter, T. Prokscha, P. M. Derlet, S. L. Lee, and L. J. Heyderman, Thermodynamic phase transitions in a frustrated magnetic metamaterial, Nat. Commun. **6**, 8278 (2015).

[4] O. Sendetskyi, L. Anghinolfi, V. Scagnoli, G. Möller, N. Leo, A. Alberca, J. Kohlbrecher, J. Lüning, U. Staub, and L. J. Heyderman, Magnetic diffuse scattering in artificial kagome spin ice, Phys. Rev. B **93**, 224413 (2016).

[5] O. Sendetskyi, V. Scagnoli, N. Leo, L. Anghinolfi, A. Alberca, J. Lüning, U. Staub, P. M. Derlet, and L. J. Heyderman, Continuous magnetic phase transition in artificial square ice, Phys. Rev. B **99**, 214430 (2019).

[6] K. Hofhuis, S. H. Skjærvø, S. Parchenko, H. Arava, Z. Luo, A. Kleibert, P. M. Derlet, and L. J. Heyderman, Real-space imaging of phase transitions in bridged artificial kagome spin ice, Nat. Phys. **18**, 699 (2022).

[7]     L. Berchialla, G. M. Macauley, F. Museur, T. Wang, A. Kleibert, P. M. Derlet, and L. J. Heyderman, *Realizing Blume-Capel Degrees of Freedom with Toroidal Moments in a Ruby Artificial Spin Ice*, arXiv:2509.01522.

[8]     E. Mengotti, L. J. Heyderman, A. F. Rodríguez, F. Nolting, R. V. Hügli, and H.-B. Braun, Real-space observation of emergent magnetic monopoles and associated Dirac strings in artificial kagome spin ice, Nat. Phys. **7**, 68 (2011).

[9]     S. Ladak, D. E. Read, G. K. Perkins, L. F. Cohen, and W. R. Branford, Direct observation of magnetic monopole defects in an artificial spin-ice system, Nat. Phys. **6**, 359 (2010).

[10]    J. C. Gartside, K. D. Stenning, A. Vanstone, H. H. Holder, D. M. Arroo, T. Dion, F. Caravelli, H. Kurebayashi, and W. R. Branford, Reconfigurable training and reservoir computing in an artificial spin-vortex ice via spin-wave fingerprinting, Nat. Nanotechnol. **17**, 460 (2022).

[11]    K. D. Stenning et al., Neuromorphic overparameterisation and few-shot learning in multilayer physical neural networks, Nat. Commun. **15**, 7377 (2024).

[12]    A. Kurenkov, J. Maes, A. Pac, G. M. Macauley, B. Van Waeyenberge, A. Hrabec, and L. J. Heyderman, Perpendicular-anisotropy artificial spin ice with spontaneous ordering: a platform for reservoir computing with flexible timescales, Commun. Eng. **4**, 183 (2025).

[13]    H. Li, L. Li, R. Xiang, W. Liu, C. Yan, Z. Tao, L. Zhang, and R. Liu, Physical reservoir computing and deep neural networks using artificial and natural noncollinear spin textures, Phys. Rev. Appl. **22**, 014027 (2024).

[14]    K. Hon, Y. Kuwabiraki, M. Goto, R. Nakatani, Y. Suzuki, and H. Nomura, Numerical simulation of artificial spin ice for reservoir computing, Appl. Phys. Express **14**, 033001 (2021).

[15]    J. H. Jensen, E. Folven, and G. Tufte, *Computation in Artificial Spin Ice*, in *The 2018 Conference on Artificial Life* (MIT Press, Tokyo, Japan, 2018), pp. 15–22.

[16]    T. Taniguchi, Echo state property and memory capacity of artificial spin ice, Sci. Rep. **15**, 9073 (2025).

[17]    H. Arava, N. Leo, D. Schildknecht, J. Cui, J. Vijayakumar, P. M. Derlet, A. Kleibert, and L. J. Heyderman, Engineering Relaxation Pathways in Building Blocks of Artificial Spin Ice for Computation, Phys. Rev. Appl. **11**, 054086 (2019).

[18]    P. Gypens, J. Leliaert, and B. Van Waeyenberge, Balanced Magnetic Logic Gates in a Kagome Spin Ice, Phys. Rev. Appl. **9**, 034004 (2018).

[19]    F. Caravelli and C. Nisoli, Logical gates embedding in artificial spin ice, New J. Phys. **22**, 103052 (2020).

[20]    Y. Li et al., Superferromagnetism and Domain-Wall Topologies in Artificial "Pinwheel" Spin Ice, ACS Nano **13**, 2213 (2019).

[21]    J. H. Jensen, A. Strømberg, I. Breivik, A. Penty, M. A. Niño, M. W. Khaliq, M. Foerster, G. Tufte, and E. Folven, Clocked dynamics in artificial spin ice, Nat. Commun. **15**, 964 (2024).

[22]     A. Farhan, P. M. Derlet, A. Kleibert, A. Balan, R. V. Chopdekar, M. Wyss, J. Perron, A. Scholl, F. Nolting, and L. J. Heyderman, Direct Observation of Thermal Relaxation in Artificial Spin Ice, Phys. Rev. Lett. **111**, 057204 (2013).

[23]     I. Gilbert, G.-W. Chern, S. Zhang, L. O'Brien, B. Fore, C. Nisoli, and P. Schiffer, Emergent ice rule and magnetic charge screening from vertex frustration in artificial spin ice, Nat. Phys. **10**, 670 (2014).

[24]     S. Zhang, I. Gilbert, C. Nisoli, G.-W. Chern, M. J. Erickson, L. O'Brien, C. Leighton, P. E. Lammert, V. H. Crespi, and P. Schiffer, Crystallites of magnetic charges in artificial spin ice, Nature **500**, 553 (2013).

[25]     P. Gypens, N. Leo, M. Menniti, P. Vavassori, and J. Leliaert, Thermoplasmonic Nanomagnetic Logic Gates, Phys. Rev. Appl. **18**, 024014 (2022).

[26]     J. C. Gartside, D. M. Arroo, D. M. Burn, V. L. Bemmer, A. Moskalenko, L. F. Cohen, and W. R. Branford, Realization of ground state in artificial kagome spin ice via topological defect-driven magnetic writing, Nat. Nanotechnol. **13**, 53 (2018).

[27]     S. Fukami, T. Anekawa, C. Zhang, and H. Ohno, A spin–orbit torque switching scheme with collinear magnetic easy axis and current configuration, Nat. Nanotechnol. **11**, 621 (2016).

[28]     L. Liu, O. J. Lee, T. J. Gudmundsen, D. C. Ralph, and R. A. Buhrman, Current-Induced Switching of Perpendicularly Magnetized Magnetic Layers Using Spin Torque from the Spin Hall Effect, Phys. Rev. Lett. **109**, 096602 (2012).

[29]     M. Cubukcu, O. Boulle, M. Drouard, K. Garello, C. Onur Avci, I. Mihai Miron, J. Langer, B. Ocker, P. Gambardella, and G. Gaudin, Spin-orbit torque magnetization switching of a three-terminal perpendicular magnetic tunnel junction, Appl. Phys. Lett. **104**, 042406 (2014).

[30]     I. M. Miron, K. Garello, G. Gaudin, P.-J. Zermatten, M. V. Costache, S. Auffret, S. Bandiera, B. Rodmacq, A. Schuhl, and P. Gambardella, Perpendicular switching of a single ferromagnetic layer induced by in-plane current injection, Nature **476**, 189 (2011).

[31]     L. Liu, C.-F. Pai, Y. Li, H. W. Tseng, D. C. Ralph, and R. A. Buhrman, Spin-Torque Switching with the Giant Spin Hall Effect of Tantalum, Science **336**, 555 (2012).

[32]     K. Garello, I. M. Miron, C. O. Avci, F. Freimuth, Y. Mokrousov, S. Blügel, S. Auffret, O. Boulle, G. Gaudin, and P. Gambardella, Symmetry and magnitude of spin–orbit torques in ferromagnetic heterostructures, Nat. Nanotechnol. **8**, 587 (2013).

[33]     X. Yin, L. Wei, P. Liu, J. Yang, P. Zhang, J. Peng, F. Huang, R. Liu, J. Du, and Y. Pu, *Field-Free Spin-Orbit Torque-Induced Switching of Perpendicular Magnetization at Room Temperature in WTe2/Ferromagnet Heterostructures*, https://arxiv.org/abs/2212.14281v1.

[34]     M. I. Dyakonov and V. I. Perel, Current-induced spin orientation of electrons in semiconductors, Phys. Lett. A **35**, 459 (1971).

[35]     A. Hoffmann, Spin Hall Effects in Metals, IEEE Trans. Magn. **49**, 5172 (2013).

[36]     J. Sinova, S. O. Valenzuela, J. Wunderlich, C. H. Back, and T. Jungwirth, Spin Hall effects, Rev. Mod. Phys. **87**, 1213 (2015).

[37]     A. Manchon, J. Železný, I. M. Miron, T. Jungwirth, J. Sinova, A. Thiaville, K. Garello, and P. Gambardella, Current-induced spin-orbit torques in ferromagnetic and antiferromagnetic systems, Rev. Mod. Phys. **91**, 035004 (2019).

[38]     J. Kim, J. Sinha, M. Hayashi, M. Yamanouchi, S. Fukami, T. Suzuki, S. Mitani, and H. Ohno, Layer thickness dependence of the current-induced effective field vector in Ta|CoFeB|MgO, Nat. Mater. **12**, 240 (2013).

[39]     T. D. Skinner, M. Wang, A. T. Hindmarch, A. W. Rushforth, A. C. Irvine, D. Heiss, H. Kurebayashi, and A. J. Ferguson, Spin-orbit torque opposing the Oersted torque in ultrathin Co/Pt bilayers, Appl. Phys. Lett. **104**, 062401 (2014).

[40]     P. Jadaun, L. F. Register, and S. K. Banerjee, The microscopic origin of DMI in magnetic bilayers and prediction of giant DMI in new bilayers, Npj Comput. Mater. **6**, 88 (2020).

[41]     Y.-T. Liu, C.-C. Huang, K.-H. Chen, Y.-H. Huang, C.-C. Tsai, T.-Y. Chang, and C.-F. Pai, Anatomy of Type-x Spin-Orbit-Torque Switching, Phys. Rev. Appl. **16**, 024021 (2021).

[42]     F. Hellman et al., Interface-induced phenomena in magnetism, Rev. Mod. Phys. **89**, 025006 (2017).

[43]     K. L. Metlov and K. Y. Guslienko, Stability of magnetic vortex in soft magnetic nano-sized circular cylinder, J. Magn. Magn. Mater. **242–245**, 1015 (2002).

[44]     T. Shinjo, T. Okuno, R. Hassdorf, † K. Shigeto, and T. Ono, Magnetic Vortex Core Observation in Circular Dots of Permalloy, Science **289**, 930 (2000).

[45]     E. C. Stoner and E. P. Wohlfarth, A Mechanism of Magnetic Hysteresis in Heterogeneous Alloys, Philos. Trans. R. Soc. Lond. Ser. A **240**, 599 (1948).

[46]     C. Phatak, A. K. Petford-Long, O. Heinonen, M. Tanase, and M. De Graef, Nanoscale structure of the magnetic induction at monopole defects in artificial spin-ice lattices, Phys. Rev. B **83**, 174431 (2011).

[47]     H. Arava, E. Y. Vedmedenko, J. Cui, J. Vijayakumar, A. Kleibert, and L. J. Heyderman, Control of emergent magnetic monopole currents in artificial spin ice, Phys. Rev. B **102**, 144413 (2020).

[48]     J. P. Morgan, A. Stein, S. Langridge, and C. H. Marrows, Thermal ground-state ordering and elementary excitations in artificial magnetic square ice, Nat. Phys. **7**, 75 (2011).

[49]     R. V. Hügli, G. Duff, B. O'Conchuir, E. Mengotti, A. F. Rodríguez, F. Nolting, L. J. Heyderman, and H. B. Braun, Artificial kagome spin ice: dimensional reduction, avalanche control and emergent magnetic monopoles, Philos. Trans. R. Soc. Math. Phys. Eng. Sci. **370**, 5767 (2012).

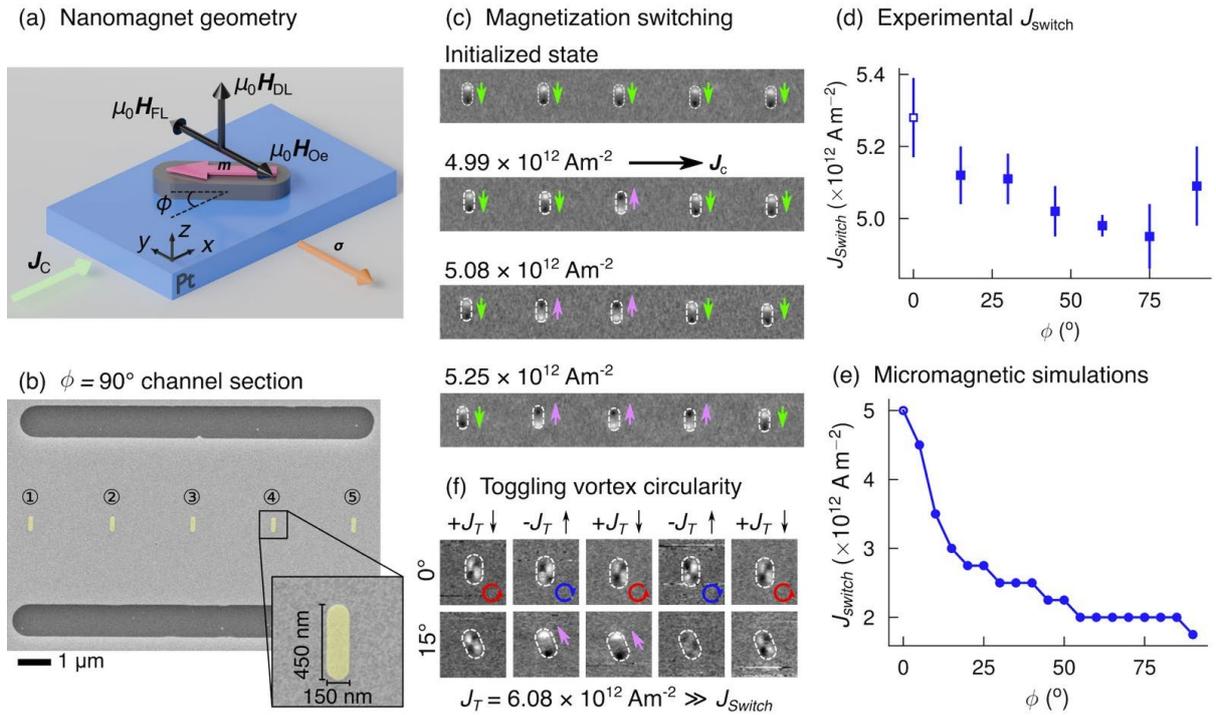

**Figure 1 | Current-induced switching of isolated nanomagnets.** (a) Schematic of an isolated stadium-shaped nanomagnet, shown in grey, with magnetization $m$ (pink arrow) on top of a heavy metal channel, shown in blue. An applied current density $J_C$ (green arrow) gives rise to the injection of spins, with spin polarization $\sigma$ pointing along the negative $y$-axis (orange arrow). This polarized spin current gives rise to field-like and damping-like effective magnetic fields ($\mu_0 H_{FL}$ and $\mu_0 H_{DL}$). The Oersted field $\mu_0 H_{Oe}$ generated by the charge current passing through heavy metal layer opposes $\mu_0 H_{FL}$. (b) SEM image of a row of nanomagnets with their long axis orthogonal to the current direction, $\phi = 90°$. (c) MFM images demonstrating the current-induced switching of single nanomagnets. The nanomagnets were first initialized with a magnetic field so that their magnetization points in the same direction given by the green arrows before 1-μs-long single current pulses were applied. As the current density of the pulses was increased, the number of switched nanomagnets increased, as indicated by the purple arrows. (d) Dependence of $\phi$ on $J_{switch}$ at which magnetization reversal occurred. The data were obtained by averaging the current density at which the middle three nanomagnets switch for each angle, with the error bars corresponding to the standard deviation. The data point at $\phi = 0°$ is kept empty to differentiate it from the rest of the data set. This is because, when current is injected, the magnetization is forced to point along the nanomagnet short axis. After the current pulse, the magnetization will relax to point along one of the two directions parallel to the long axis with no directional preference, so that it is equally likely for the nanomagnet to return to its initial magnetization state as it is to switch. Therefore, we cannot define the $J_{switch}$ for this angle in the way we did it for the rest of the angles. Nevertheless, we have still indicated the value at which a reversal in the magnetic state was observed. (e) Simulated angular dependence of a single nanomagnet using micromagnetic simulations. (f) MFM images demonstrating the ability to control the vortex circularity of larger nanomagnets in the Type $x$ geometry where $\phi = 0°$. These vortex states are not well controlled when the angle is changed to 15° where several different magnetic states can be seen following application of the current pulse.

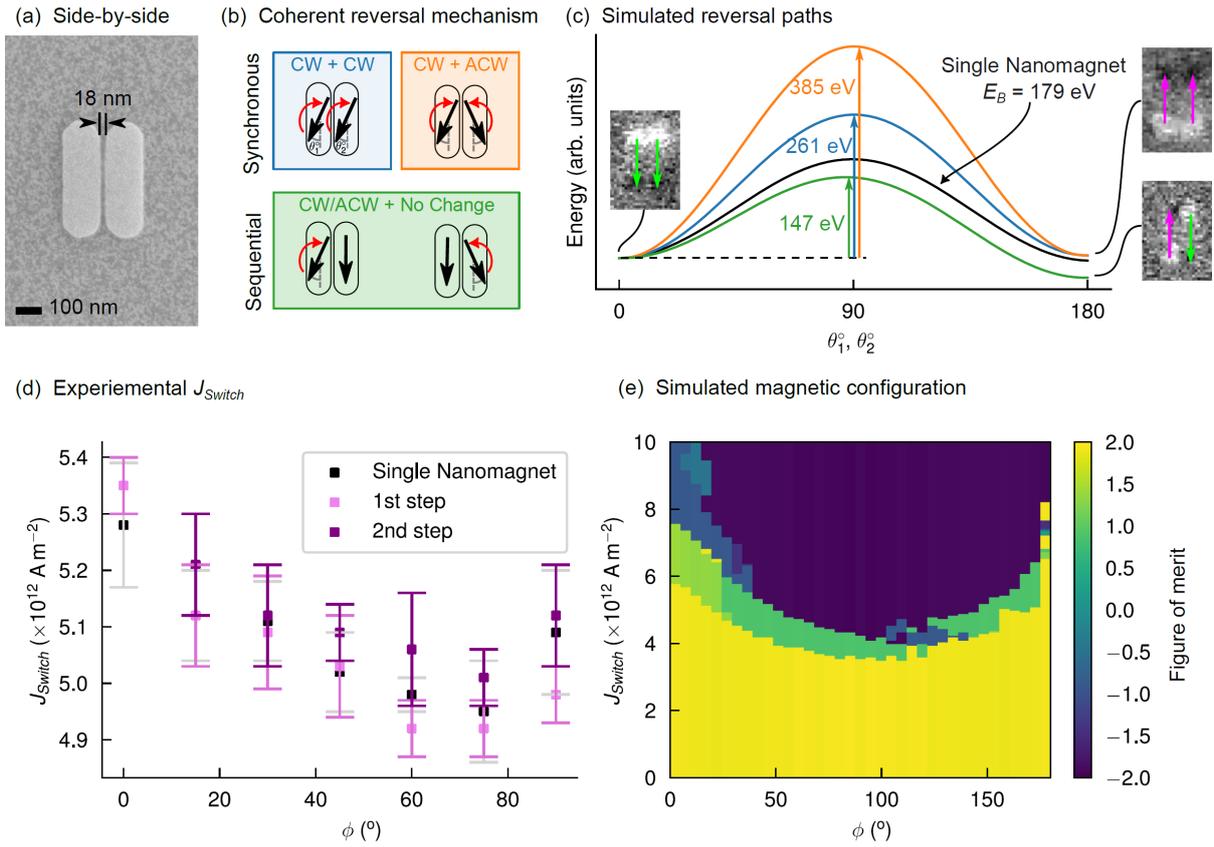

**Figure 2 | Current-induced switching of side-by-side pairs of nanomagnets.** (a) SEM image of a side-by-side nanomagnet pair. (b) Schematic of the coherent rotation mechanism of the magnetization in the two nanomagnets. For synchronous reversal, the magnetization in the two nanomagnets rotates simultaneously, either in the same sense or in opposite senses. For sequential reversal, one of the nanomagnets switches first by clockwise or anticlockwise rotation of the magnetization. (c) Reversal paths with energy barriers for magnetization switching of the side-by-side pair of nanomagnets determined from micromagnetic simulations based on the coherent reversal shown in part (b). The reversal path is color-coded in green for sequential switching, and blue and orange for the two types of synchronous switching. On either side of the energy barrier, MFM images of the experimentally observed magnetic configurations are shown. (d) Experimental dependence of $J_{switch}$ on $\phi$ for side-by-side nanomagnetic pairs. Here, we observe a sequential switching process, first switching from a parallel to an energetically favorable antiparallel configuration. On further increasing the current density, we then observe switching from an antiparallel configuration to a parallel configuration with the magnetization in nanomagnets reversed relative to the initial state. The minimum switching current $J_{switch}$ for the 1st and 2nd steps are indicated with light and dark purple data points, respectively. These are taken as the average threshold switching current for the three central nanomagnet pairs on the channel at positions ②-④, with the error bars corresponding to the standard deviation. The data for isolated nanomagnets is shown with black markers and grey error bars for comparison. (e) Simulated magnetic configuration of side-by-side nanomagnet pairs following the injection of an electrical current pulse where the figure of merit is given by $m1_{start} \cdot m1_{end} + m2_{start} \cdot m2_{end}$, highlighting the threshold current required to switch nanomagnets for a given $\phi$. For the initial magnetic configuration, the figure of merit is +2 (yellow region) with the magnetization in both nanomagnets $m$ = -1. When the figure of merit is -2 (dark blue region), it represents the opposite parallel magnetic configuration and, between these two regions, the region in green has a figure of merit of 0, where at least one nanomagnet has switched giving an antiparallel configuration.

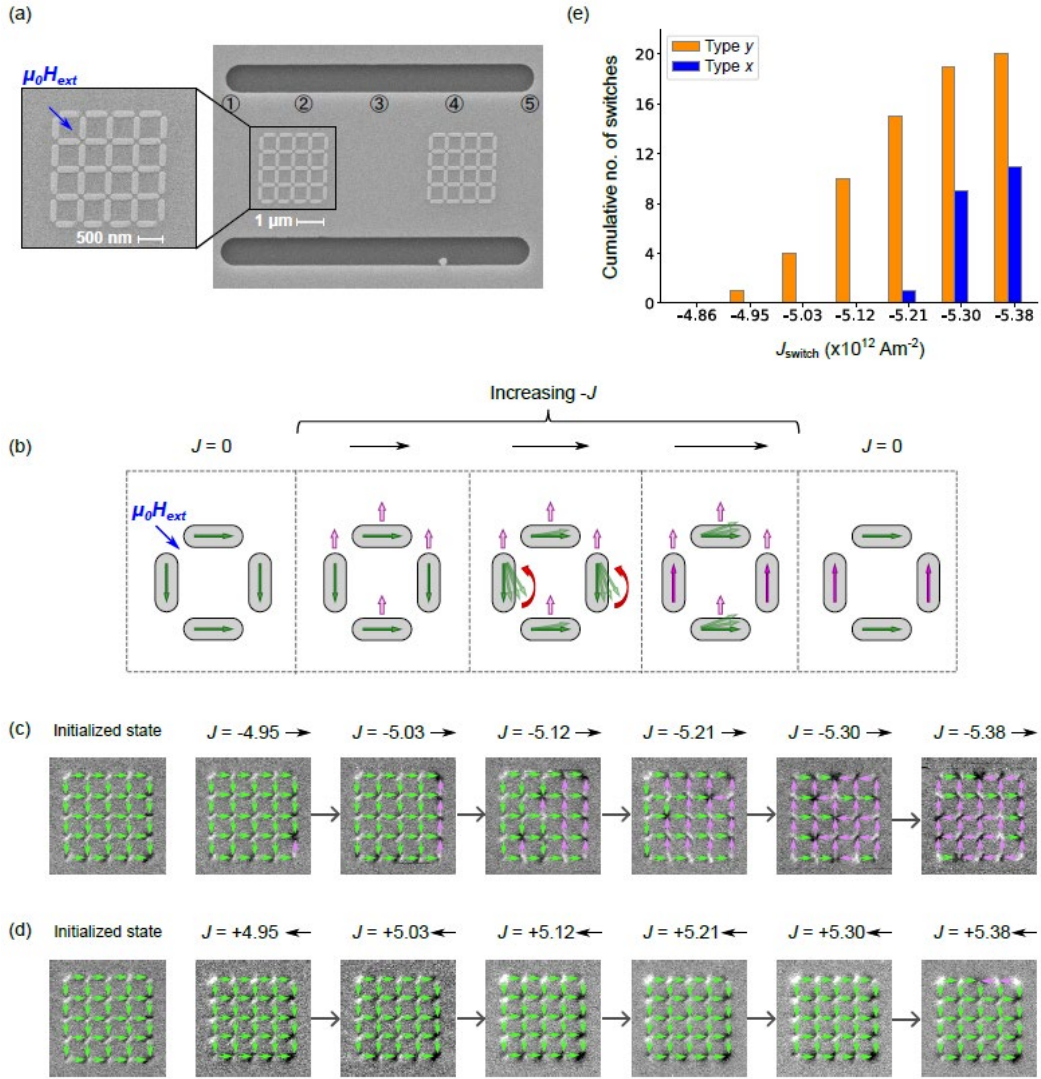

**Figure 3 | Current-induced switching of nanomagnets within an artificial square ice.** (a) SEM image of an artificial square ice with the direction of the applied magnetic field $\mu_0 H_\text{ext}$ shown with a blue arrow. This is one of two 4 × 4 arrays of nanomagnets placed on the channel, and is located in position ②. The numbers shown at the top indicate the locations of the single nanomagnets in Fig. 1(b). (b) Schematic of the expected switching behavior in a single square ring of nanomagnets. As shown in the leftmost schematic, the system is first initialized so that the magnetization in the vertical (horizontal) nanomagnets points down (to the right). A current of negative polarity applied in the horizontal direction generates an effective field in the vertical direction, indicated by the purple arrows. This effective field promotes rotation of the magnetization within nanomagnets. For a sufficiently large current density, the magnetization in the vertical nanomagnets is switched, while it is not large enough to switch the magnetization in the horizontal magnets. When the current is removed (rightmost schematic), the system relaxes with the vertical nanomagnets switched. (c) MFM images of the magnetic state after each current pulse, with the current density (×$10^{12}$ Am$^{-2}$) shown above each image. Starting with the initialized state shown on the left, with all arrows in green, a negative-polarity current pulse is applied horizontally. The macrospins of nanomagnets that have switched are indicated with purple arrows. (d) The measurement is repeated, but this time the polarity of the applied current is positive. (e) Cumulative number of switched nanomagnets as a function of increasing current density for the same 4 × 4 array of square rings of nanomagnets. Here, Type $y$ (Type $x$) refers to the nanomagnets whose long axis is oriented along the vertical (horizontal) direction, perpendicular (parallel) to the current direction.

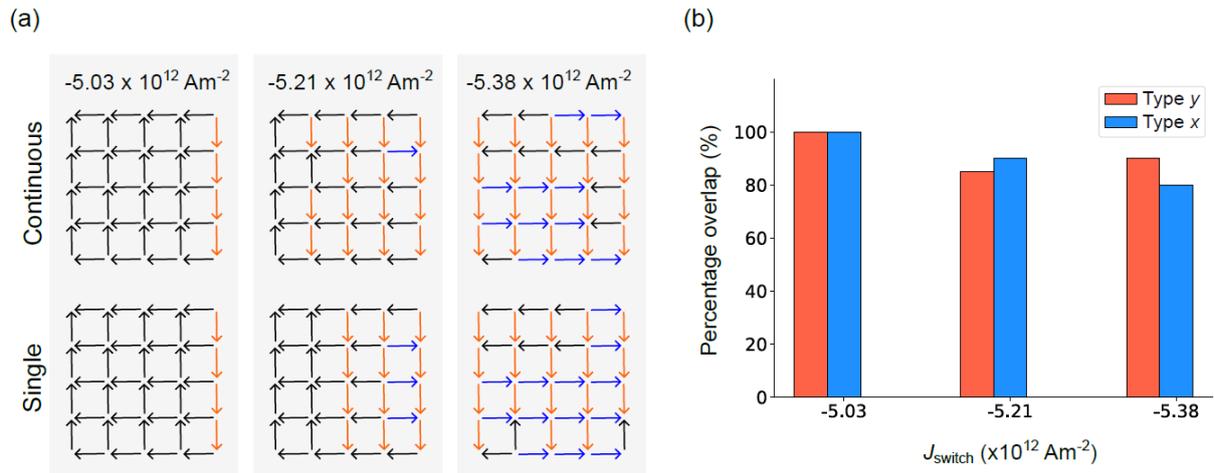

**Figure 4 | Comparing switching in an artificial square ice with a single current pulse and a series of pulses with increasing current density.** (a) Schematic of the experimentally measured magnetic configuration after applying current pulses in two different ways. Top row: The magnetic state is shown after current pulses are applied with increasing current density without re-initialization in a magnetic field. The first pulse is applied with a current density of -4.86×10$^{12}$ Am$^{-2}$. For subsequent current pulses, current was then stepped down to even more negative values in steps of -0.09×10$^{12}$ Am$^{-2}$ steps, although we only show the state after selected current pulses. Lower row: each state is obtained after a combination of first initializing the sample and then applying a single pulse of the stated current density. (b) Graph showing the percentage of overlap, i.e., the percentage of nanomagnets with the same magnetization direction when implementing the applied current schemes shown in (a) for a given current density for both Type *y* and Type *x* nanomagnets.

**Supplemental Materials for "Current-induced magnetization control in dipolar-coupled nanomagnet pairs and artificial spin ice"**

A. Pac[1,2*], G. M. Macauley[1,2†], J. A. Brock[1,2], A. Hrabec[1,2], A. Kurenkov[1,2], V. Raposo[3,4], E. Martinez[3,4], and L. J. Heyderman[1,2*]

1. Laboratory for Mesoscopic Systems, Department of Materials, ETH Zurich, 8093 Zurich, Switzerland
2. PSI Center for Neutron and Muon Sciences, 5232 Villigen PSI, Switzerland
3. Department of Applied Physics, University of Salamanca, 37008 Salamanca, Spain
4. Unidad de Excelencia en Luz y Materia Estructuradas (LUMES), Universidad de Salamanca, Salamanca, Spain

Corresponding authors: aleksandra.pac@psi.ch, laura.heyderman@psi.ch

† Present Address: Department of Physics, Princeton University, Princeton, NJ 08540 USA

**Section 1: Channel design and current density distribution**

In this section, we discuss the rationale behind the conducting channel design in relation to the current density distribution, as determined from finite element simulations performed using COMSOL. The primary objective of the channel design was to accommodate multiple identical nanomagnets along the channel, ensuring that the orientation of the nanomagnets with respect to the current, given by the angle $\phi$, was the same for all nanomagnets. This arrangement allowed us to measure the threshold current density required to switch the nanomagnets ($J_{switch}$) for several nanomagnets simultaneously, enabling the determination of the average threshold current density and its standard deviation.

**Section 1.1: Channel layout**

In Figure S1.1, SEM images of a channel device at various magnifications are presented. The channel label, shown in Figure S1.1(a), aided the location of specific devices on the chip. The spacing between individual elements (e.g., nanomagnets and pairs of nanomagnets) was chosen to be sufficient to assume that the dipolar interactions between them were negligible and did not affect the switching current. For example, as shown in Fig. S1.1 (b), the separation between the isolated nanomagnets was ~1.5 µm. Additionally, the incorporation of multiple horizontal slits within the channel [see Fig. S1.1 (b)] ensured a more uniform current distribution in the areas between the slits when current was applied. Accordingly, the nanomagnets were placed in the regions between the slits, with positions labelled ①-⑤ in Fig. S1.1(b). The uniformity of the current density was verified using COMSOL simulations and is shown in Fig. S1.2 (a). Lastly, to maximize the strength of dipolar interaction between the nanomagnets in pairs of nanomagnets and artificial spin ices, the separation between them was kept to the minimum achievable with nanofabrication. The minimum edge-to-edge separation between two end-to-end nanomagnets was ~22 nm, as shown in Fig. S1.1(c), while for the side-by-side pairs the separation was ~18 nm [see Fig. 2(a) in the main text]. The separation between nanomagnets within the artificial square ice was ~26 nm.

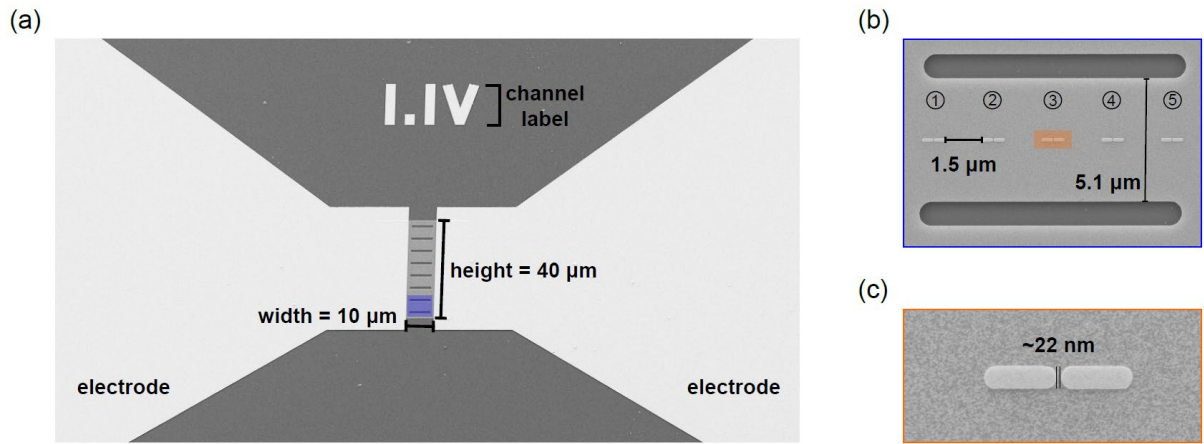

**Figure S1.1 | Channel layout used for single nanomagnets, pairs of nanomagnets and artificial spin ices.** (a) Layout of the channel, here featuring end-to-end pairs of nanomagnets. All channels have the same design. The channel label is made from a Cr (5 nm)/Au (100 nm) bilayer, which is also used for the electrodes on both sides of the channel. The channel width is 10 μm and height is 40 μm. Eight horizontal slits are incorporated to promote a homogeneous current distribution within the channel. (b) Magnified view of one of the seven channel sections between the slits on which the nanomagnets are placed, in this case five pairs of end-to-end nanomagnets. The separation between the nanomagnet pairs is approximately 1.5 μm, which is sufficient to prevent dipolar interactions between the different pairs. The vertical edge-to-edge separation between adjacent slits is ~5.1 μm. The nanomagnets are positioned at the midpoint between slits. (c) For the end-to-end pair of nanomagnets, the nanomagnet separation is minimized to make dipolar interactions as large as possible, with an edge-to-edge spacing of ~22 nm.

**Section 1.2: Simulated current density distribution within the channel**

In Figure S1.2, the current density distribution within the channel is shown, which was simulated using COMSOL finite element analyzer, including the regions near the slits and the channel sections where the nanomagnets are placed. The simulations are carried out with a square current pulse with an applied current density of $5.21 \times 10^{12}$ Am$^{-2}$ and pulse width of 1500 ns. The state observed at 1400 ns after the start of the (square) current pulse is shown in Fig. S1.2(a). From at the increase of the current density as a function of time, we find that we reach steady-state current density within the channel at 1400 ns, which allows us to determine the current density distribution at different locations. At positions ②, ③ and ④, corresponding to the nanomagnets in the three central positions on the channel, the current density is relatively uniform. However, there are significant spatial variations in the current density near the horizontal edges of the slits, which results in relatively large variations in the current density at locations corresponding to nanomagnets ① and ⑤. Therefore, nanomagnets at locations ① and ⑤ were excluded from our data analysis. The variations in current density along the channels are illustrated in Fig. S1.2(b). Here we plot the current density between the slits at locations ① - ⑤ where the nanomagnets are positioned for each of the channel sections, which are labelled R1-R7 in Fig. S1.1.

For the experiments, the current pulse width was kept constant at 1000 ns. Using an oscilloscope, we confirmed that the shape of the pulse is square and, as for the COMSOL simulations, we assume that we reach a steady state in the current density.

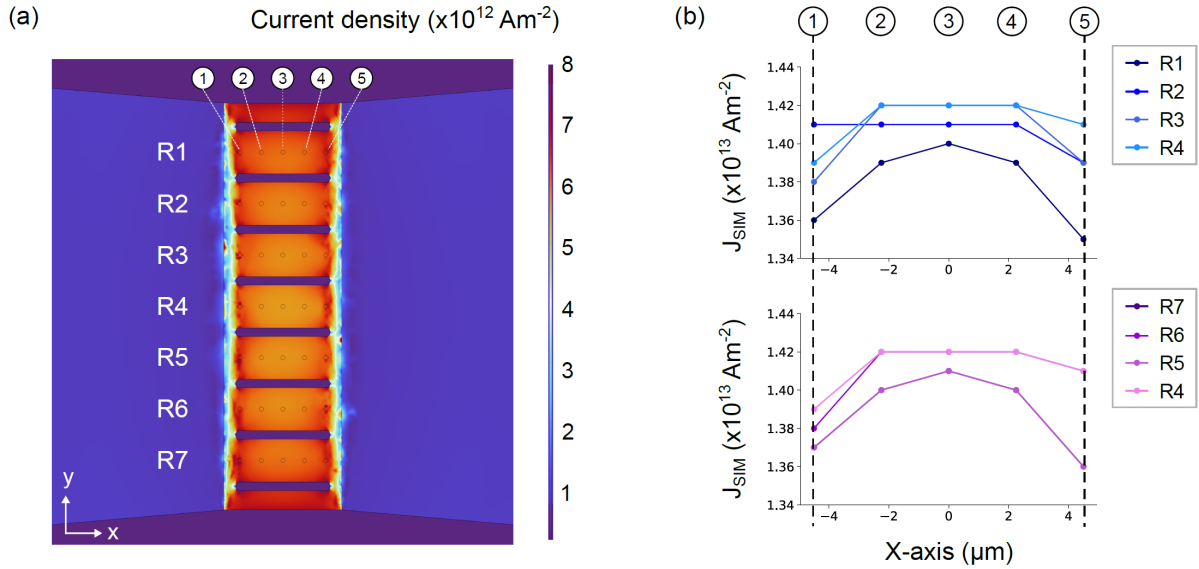

**Figure S1.2 | Current density distribution within the channel.** (a) Current density distribution obtained from COMSOL simulations. The whole channel is 10 µm long and 40 µm wide as in the experiment (see Fig. S1.1). The numbers in white discs, ①-⑤, indicate the locations of the nanomagnets within the channel using the convention established in the main text. The labels R1-R7 refer to the horizontal channel sections, where each channel section corresponds to different orientation of the nanomagnets. (b) Current density values plotted for each nanomagnet position, separated into the upper half of the channel with sections R1-R4 (upper panel) and the lower half of the channel with sections R4-R7 (lower panel). Due to the symmetry of the channel design, the current distributions in the upper and lower panels are expected to be similar. The positions of the nanomagnets (along the x-axis) are labelled above the plots. Dashed lines correspond to positions ① and ⑤, where a reduction in current density is observed compared to positions ②–④. The maximum deviation of the current densities is <4% from the highest value that is found in the middle of the channel. Additionally, there is a decrease in current density near nanomagnets ① and ⑤, which are close to the edges of the slits, and this is also evident in part (a).

## Section 2: Estimate for the Oersted field generated by the current passing through the heavy metal layer

In this section, we examine the role of the Oersted field in determining the threshold current density ($J_{switch}$) required for switching of isolated nanomagnets. Using COMSOL simulations, we calculate the approximate *y*-component of the Oersted field and analyze how it varies as a function of the nanomagnet thickness. From this, we extract an Oersted field coefficient, $b = \frac{\Delta B}{\Delta J} = 3.8$ (1mT/$10^{12}$ Am$^{-2}$), as shown in Fig. S2.2, where $b$ is the proportionality constant that was obtained by linearly fitting the points of current density vs the value of the Oersted field. $J$ represents the current density that was applied to the system from which corresponding Oersted field was obtained. Note that we use interchangeably $B$ or $\mu_0 H$ in this document.

# Section 2.1: Oersted field estimate for a single current pulse

(a) Top view of the y-component of magnetic flux

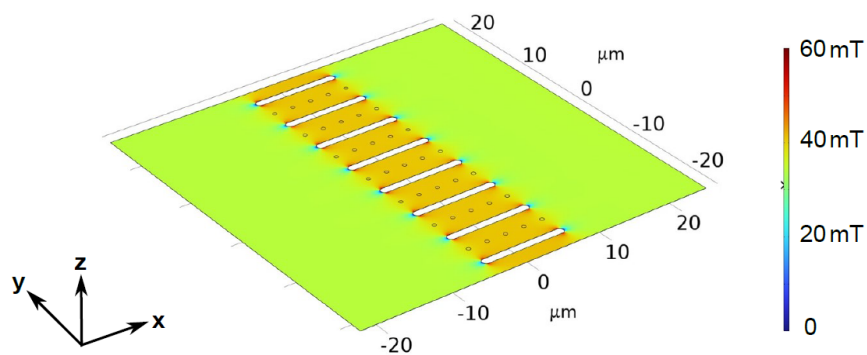

(b) Side view of the y-component of magnetic flux

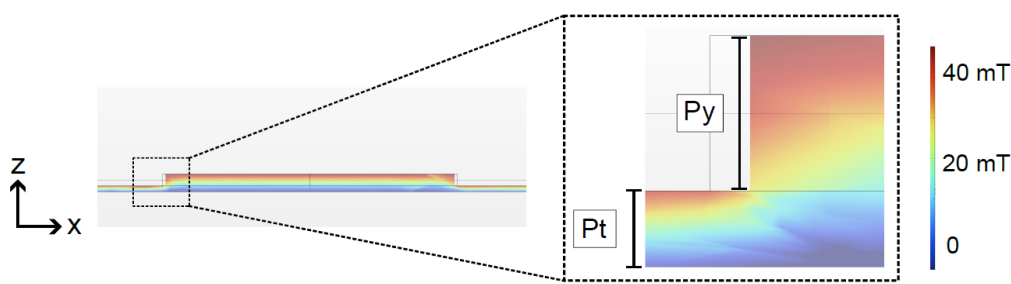

**Figure S2.1 | y-component of the simulated Oersted field in the channel.** (a) Three-dimensional top view of the simulated channel showing the distribution of the y-component of the Oersted field at the top surface of the channel. The applied current density is $5.21 \times 10^{12}$ Am$^{-2}$ along the negative x-axis direction. In the channel sections, between the horizontal slits, the Oersted field is predominantly uniform. (b) Side view of a Py nanomagnet on top of the Pt channel illustrating the variation of the y-component of the Oersted field through the thickness. At the top surface of the nanomagnet, the y-component of the Oersted field reaches approximately 40 mT, decreasing to ~25 mT at the center of the nanomagnet.

**Section 2.2: Oersted field as a function of current density**

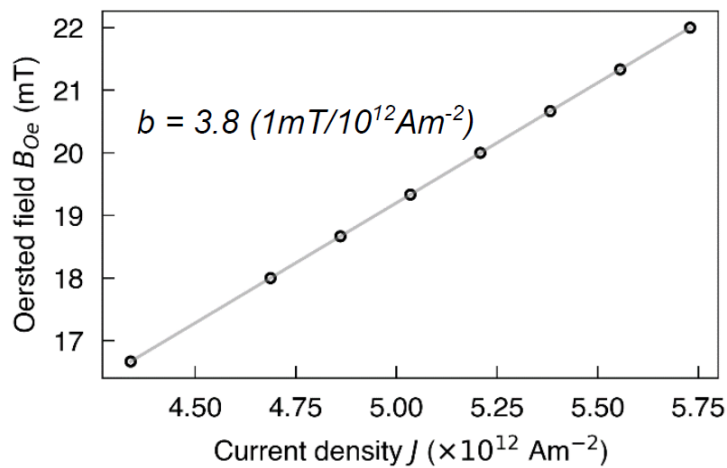

**Figure S2.2 | Dependence of the Oersted field at the center of a nanomagnet as a function of the current density $J$, based on the geometry of the COMSOL simulations shown in Fig. S2.1**. A linear fit (grey line) to the data (black markers) yields a proportionality constant—the Oersted field coefficient—of $b = 3.8$ (1mT/$10^{12}$ Am$^{-2}$). This value is used to model the expected strength of the Oersted field in our micromagnetic simulations.

**Section 3: Estimates of the Switching Fields and Energy Barrier to Switching of Isolated and Coupled Nanomagnets**

In this Sections 3.1 and 3.2, we report two ways in which we have estimated the energy barrier to switching of isolated and coupled nanomagnets: (i) by performing hysteresis loop simulations to find their coercive fields, and (ii) by modelling a coherent rotation of the magnetization within nanomagnets. These two complementary methods allow us to estimate the magnitude of the effective fields generated by spin-orbit torques that are necessary to overcome the energy barrier. These simulations also help to explain some features of the angular dependence of the threshold current switching curves shown in Figs. 1 and 2 of the main text. From our initial simulations we observe that the sequential switching in the end-to-end nanomagnet pairs should be more favorable due to the lower energy barrier. In the experimental results shown in Section 4, however, we note that sequential switching is never seen. This discrepancy can most likely be attributed to the fact that the step sizes used for the current density in the experiment were too large to capture the sequential switching.

We performed micromagnetic simulations using MuMax3 [1]. For these, we use a cell size of 2.5 nm x 2.5 nm x 2.5 nm, an exchange constant of 13 pJ/m, and a saturation magnetization of 860 kA/m. Since MuMax3 is a finite-difference micromagnetic solver, it discretizes space into cuboid-shaped cells. This approach can lead to inaccuracies when modeling structures with curved edges, such as at the ends of our nanomagnets, because the cuboid grid cannot perfectly represent smooth curves. To achieve smoother geometries, we set an edge smoothing level of 8 in MuMax3, which subdivides each simulation cell into $8^3$ = 512 subcells. The saturation magnetization of cells at the edge of the nanomagnet volume is then adjusted based on how many subcells fall within boundary of the defined shape. For the hysteresis loop simulations, we use a damping constant of 0.5 to accelerate the convergence of magnetization dynamics towards the equilibrium, and the smallest field step close to the coercive field is 0.25 mT.

**Section 3.1: Hysteresis Loop Simulations to Determine the Coercive Field for an Isolated Nanomagnet**

In Fig. S3.1(a), we show simulated magnetization-field (*M-B)* loops of an isolated stadium-shaped nanomagnet (*l* = 450 nm, *w* = 150 nm, *t* = 10 nm) with the magnetic field applied parallel (orthogonal) to the long axis of the nanomagnet given by the blue (orange) curve. The geometry for the simulation is shown in the inset, with the long and short axes of the nanomagnet are aligned with the *x*- and *y*- axes of the simulation grid, respectively. The applied magnetic field angle α is with respect to the long axis of the nanomagnet. Here, we define the coercive field $H_C$ to be the magnitude of the magnetic field at which the magnetization along the nanomagnet long axis, i.e. $M_x$, changes sign. In Fig. S3.1(a), it can be seen that the coercive field is larger when the magnetic field is applied along the short axis than when it is applied along the long axis.

The hysteresis loop is simulated over a range of intermediate applied field angles, to obtain the switching field $H_C(α)$ for an isolated nanomagnet, which follows approximately the Stoner-Wohlfarth model. This is shown in Fig. S3.1(b), where the left-hand axis denotes the coercive field in mT. On the right-hand axis of Fig. S3.1(b), the coercive field is converted into an estimate for the energy barrier of the nanomagnet $E_B = M_S V H_C$, where $M_S$ and $V$ are the saturation magnetization and volume of the nanomagnet, respectively.

In Fig. S3.1(b), we observe that, as the angle of the applied magnetic field is increased from 0º, the coercive field initially decreases, reaching a minimum of approximately 21 mT at α = 22º. Beyond this angle, the coercive field steadily increases to 79 mT at α = 89º, where the applied magnetic field is nearly perpendicular to the long axis of the nanomagnet. Our definition of the switching field assumes that there is always a component of the magnetization **M** along the long axis of the nanomagnet, so along *x*, with switching corresponding to the change in sign of this component $M_x$. This assumption is no longer appropriate when the field is applied along the short axis, i.e. when α = 90º. In this case, a sufficiently strong field causes the magnetization within the nanomagnet to align primarily along the short axis. As a result, there is no significant component of **M** along the long axis. Therefore, no value for the switching field is given for α = 90º in Fig. S3.1(b).

Because of the geometric symmetry of the nanomagnet, the coercive field asteroid is symmetrical about α = 90º, i.e. $H_C(α) = H_C(180° - α)$. The minimum coercive field $H_C$ occurs at α = 22º and α = 158º. These angles are indicated with dashed vertical lines in Fig. S3.1(b). Similar to the trend in the switching current as a function of angle, with a minimum at a specific angle (main text, Fig. 2), we observe a minimum in $H_C$ at $α ≠ 0°$. For SOT-driven switching, the minimum switching current for an isolated nanomagnet occurs when the nanomagnet is orientated at 15º to the current direction. The discrepancy between the experimentally observed angle in a current (15°) and the simulated value in a magnetic field (22°) may be attributable to the coarse angular resolution used in the experiments. Specifically, experimental measurements were conducted with angular step sizes of 15° from 0° to 90°. Consequently, the experimentally reported value of 15° represents the closest available measurement point to the simulated result of 22°.

We also determined the effect of interfacial DMI (iDMI) due to the Pt/Py interface on the coercive field, and we found that an iDMI of 1 mJ/m$^2$ slightly reduces the switching field at every applied field angle by approximately 2 mT.

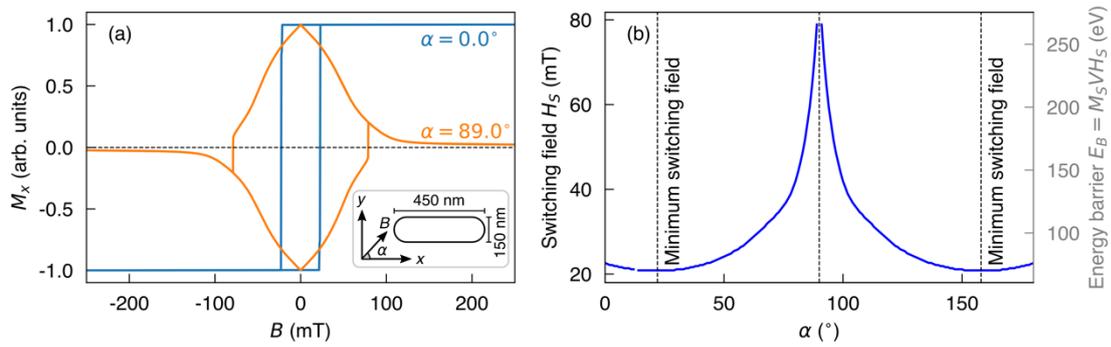

**Figure S3.1 | Micromagnetic simulations of the magnetization reversal of an isolated nanomagnet**. (a) Hysteresis loops (magnetization along x, $M_x$, versus magnetic field, $B$) of an isolated nanomagnet with the magnetic field applied along its long (short) axis in blue (orange). The inset shows the orientation of the nanomagnet with respect to the x-and y-axes of the simulation grid, as well as the definition of α, the angle of the applied magnetic field with respect to the long axis of the nanomagnet. The dimensions of the nanomagnet are 450 nm x 150 nm x 10 nm, matching those used in the experiment. The coercive field, which is a measure of the energy barrier to reversal of the nanomagnet, is taken as the magnitude of the applied magnetic field $B$ that reverses the component of $M$ along the long axis of the nanomagnet $M_x$, so at $M_x = 0$. (b) The dependence of the coercive field $H_C$ on α is obtained by simulating the magnetization reversal for a range of angles. Local minima in $H_C$ correspond to the field applied along α = 22° and α = 158° to the long axes of the nanomagnet. At α = 90°, any switching of the magnetization occurs along the short axis and not the long axis of the nanomagnet. As a result, our definition of the switching field, which relies on $M_x$ changing sign, does not occur. To reflect this, no switching field value is shown for α = 90°.

## Section 3.2: Energy barrier to switching determined by modelling coherent rotation of the magnetization within the nanomagnets

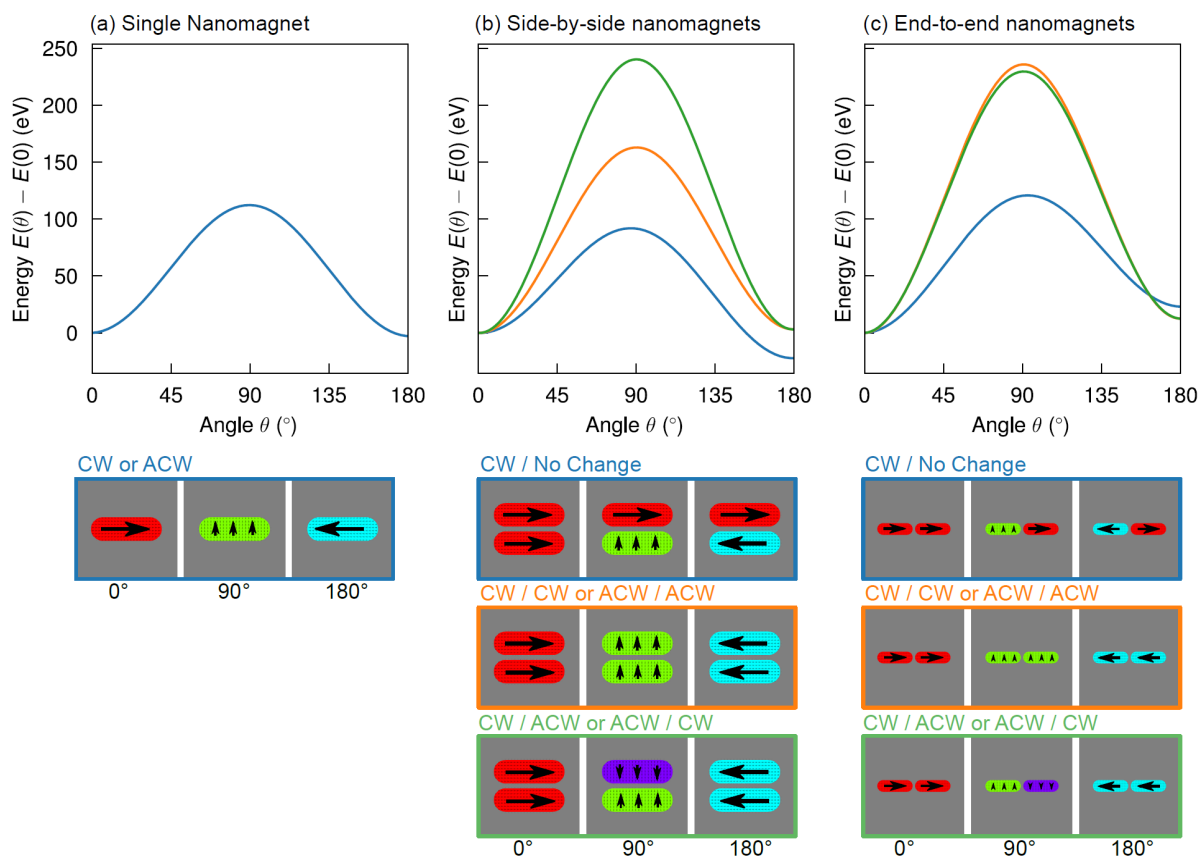

**Figure S3.2 | Energy barrier estimation using micromagnetic simulations assuming coherent magnetization reversal.** (a) For an isolated nanomagnet, magnetization reversal can occur via either clockwise (CW) or anticlockwise (ACW) rotation of the magnetization, where both paths are energetically equivalent. The initial and final states are shown in the blue box in the left and right panels, respectively, and the intermediate state, where the magnetization aligns along the hard axis at 90°, is shown in the middle panel. Here, the switching process occurs in a single step (blue curve). (b) For side-by-side nanomagnet pairs, switching can either involve a single nanomagnet or occur in both nanomagnets simultaneously. When both nanomagnets switch together (orange and green boxes and curves), the energy barrier is significantly higher than for single-nanomagnet switching (blue box and curve). This explains the experimentally observed sequential (or two-step) switching process (shown in Fig. 2(d) of the main text). (c) For end-to-end nanomagnet pairs, the energy barriers appear to favor a sequential switching (blue box and curve). However, when one of the nanomagnets reverses first, this temporarily places the pair in a higher-energy state, while the reversal of the second nanomagnet restores the system to a lower-energy state. While sequential switching of nanomagnets is occasionally observed in experiments, simultaneous switching (orange and green boxes and curves) is more commonly observed.

## Section 4: Experimental results for end-to-end pairs of nanomagnets

In this section, we present the experimental results for the end-to-end pairs of nanomagnets that, unlike the side-by-side pairs of nanomagnets, switch simultaneously via single step. Starting with the initial state where magnetization of both nanomagnets is along the same direction, when the current is injected, both nanomagnets reverse their magnetization synchronously. The average $J_{switch}$ for the middle three pairs of nanomagnets for each angle are shown in Fig. S4.1. The error bars represent the standard deviation for each angle. Here, we observe a similar trend to that observed for isolated and side-by-side pairs of nanomagnets where the minimum occurs at $\phi = 75°$ with $J_{switch} = 5.01 \times 10^{12}$ Am$^{-2}$ and maximum at $\phi = 0°$ with $J_{switch} = 5.35 \times 10^{12}$ Am$^{-2}$.

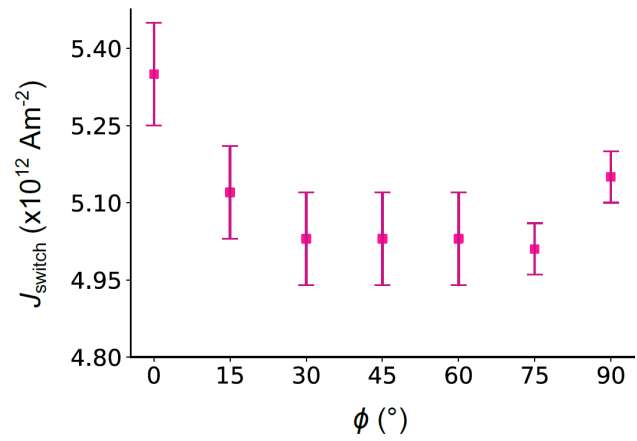

**Figure S4.1 | End-to-end nanomagnet switching.** Experimental results for the angular dependence of the threshold switching current $J_{switch}$ for end-to-end pairs of nanomagnets. The average is for the three middle pairs of nanomagnets at positions ②–④ for each angle. The error bars represent the standard deviation for each angle.

## Section 5: Single nanomagnet switching – additional scenarios

In this section, we present experimental data for two additional switching scenarios involving isolated nanomagnets. Figure S5.1 illustrates the three scenarios analyzed. In the first scenario (left panel), we show the configuration discussed in the main text, indicating the directions of all relevant components during spin-orbit torque (SOT) switching. The second scenario maintains the same initial magnetization direction, but the current is reversed. In this case, the field-like effective field component of SOT aligns with the initial magnetization, preventing switching in nanomagnets of Type *y* (with long axis at 90° to the current). In the third scenario, both the initial magnetization direction and current are reversed, creating a symmetric counterpart to the first scenario. We would therefore expect the same behavior in Scenario 3 as in Scenario 1.

The experimental threshold switching currents as a function of angle for Scenarios 1-3 are shown in Figures S5.2(a-c), respectively.

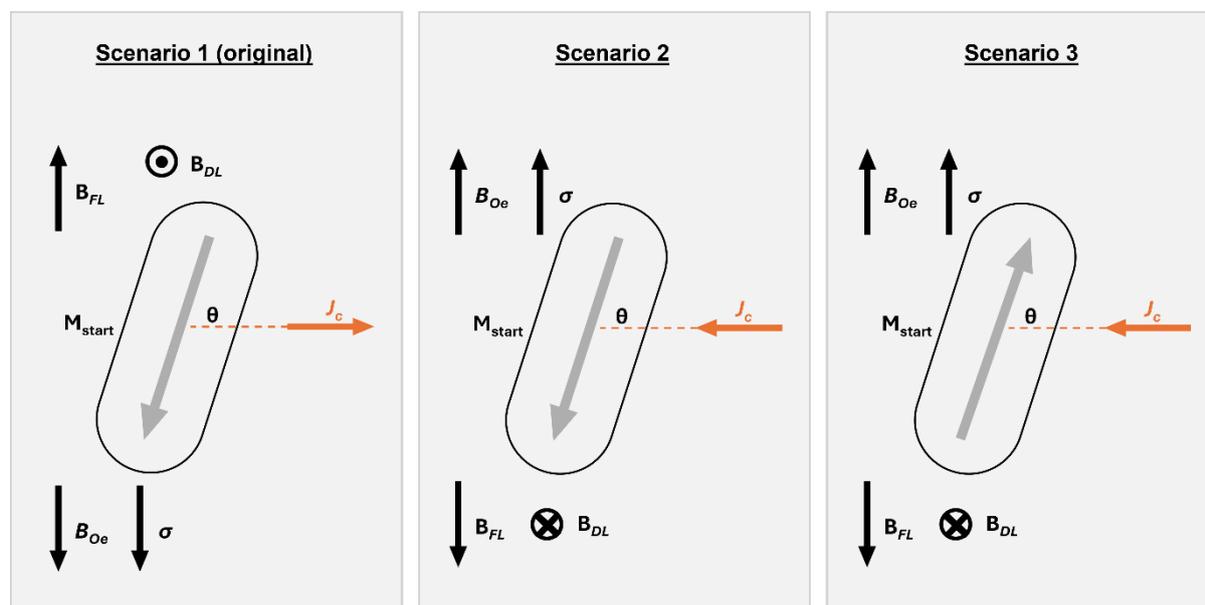

**Figure S5.1 | Three switching scenarios for an isolated nanomagnet.** Scenario 1 (left) is the original configuration that was used to obtain the angular dependence of threshold current density for an isolated nanomagnet. The direction of initial magnetization of the nanomagnet, $M_{start}$, conventional current, $J_c$, spin polarization, $\sigma$, field-like and damping-like effective fields, $B_{FL}$ and $B_{DL}$, and Oersted field, $B_{Oe}$, are indicated. Scenario 2 (middle) has the same initial magnetization direction as Scenario 1, but here the current direction is reversed. This leads to change in direction of the spin polarization, effective fields and Oersted field, making it more difficult to switch the nanomagnet. In Scenario 3 (right), both the reversed magnetization and conventional current direction are reversed with respect to Scenario 1. This means that we should expect the same switching behavior as that of Scenario 1, since Scenario 3 is a symmetric counterpart to Scenario 1.

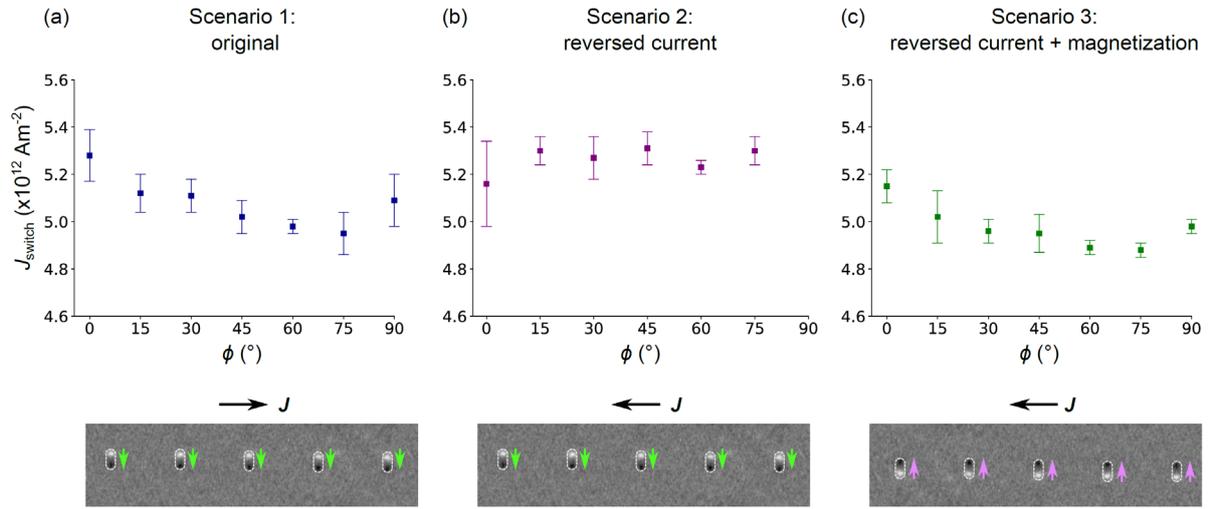

**Figure S5.2 | Threshold switching current for isolated nanomagnets for the three scenarios.** (a) Threshold current density, $J_{switch}$, required for switching of isolated nanomagnets in Scenario 1 [Fig. S5.1(a)] for different $\phi$. This is the original switching scenario and reproduced from Fig. 1(e) in the main text. Here the minimum in $J_{switch}$ is $4.95 \times 10^{12}$ Am$^{-2}$ at $\phi = 75°$. (b) The results for Scenario 2 where the initial magnetization direction is kept the same as in Scenario 1, but the current direction is reversed. Here we do not observe an obvious trend in switching and there is no distinct minimum. Additionally, we observe lack of switching for nanomagnets at $\phi = 90°$ since the direction of initial magnetization is in the direction of the $\boldsymbol{B_{FL}}$ generated by SOTs. (c) For Scenario 3, both the initial magnetization and current directions are reversed with respect to Scenario 1. This is a symmetric counterpart to Scenario 1 and thus a similar trend for $J_{switch}$ versus $\phi$ is seen, with a minimum $J_{switch}$ of $4.88 \times 10^{12}$ Am$^{-2}$ at $\phi = 75°$. Below each graph, the direction of the initial magnetization is shown on MFM images of the isolated nanomagnets, along with the direction of the current $\boldsymbol{J}$, for each scenario.

### Section 6: Micromagnetic simulations for single nanomagnet switching

To interpret the experimental results shown in Fig. 1(e), micromagnetic simulations were conducted using the following material and geometric parameters: an exchange constant of $A_{ex} = 13$ pJm$^{-1}$ and a saturation magnetization of $M_s = 860$ kAm$^{-1}$. The dimensionless Gilbert damping constant was set to 0.02, and the spin Hall angle was set to $\theta_{SH} = +0.2$. The permalloy has no uniaxial anisotropy ($K_u = 0$). The dimensions of the nanomagnet are a length l = 450 nm, width w = 150 nm, and ferromagnetic layer thickness $t_{FM} = 10$ nm, placed on top of a heavy metal layer of thickness $t_{HM} = 9$ nm. The current density $\boldsymbol{J} = J_{HM}\boldsymbol{u}_x$ is applied in the heavy metal layer, and is always directed along the x-axis. The applied current $J(t)$ is a pulse with a duration of $t_p = 8$ ns. In the simulations, the field-like torque component is assumed to be proportional to the damping-like torque, expressed as $H_{FL} = kH_{DL}$. The Oersted field is included and calculated according to $H_{Oe} = -(J_{HM}(t)t_{HM} / 2)$. The overview of the micromagnetic simulations and different components of the system are depicted in Fig. S6.1, for various arrangements of the nanomagnet with respect to the current polarity.

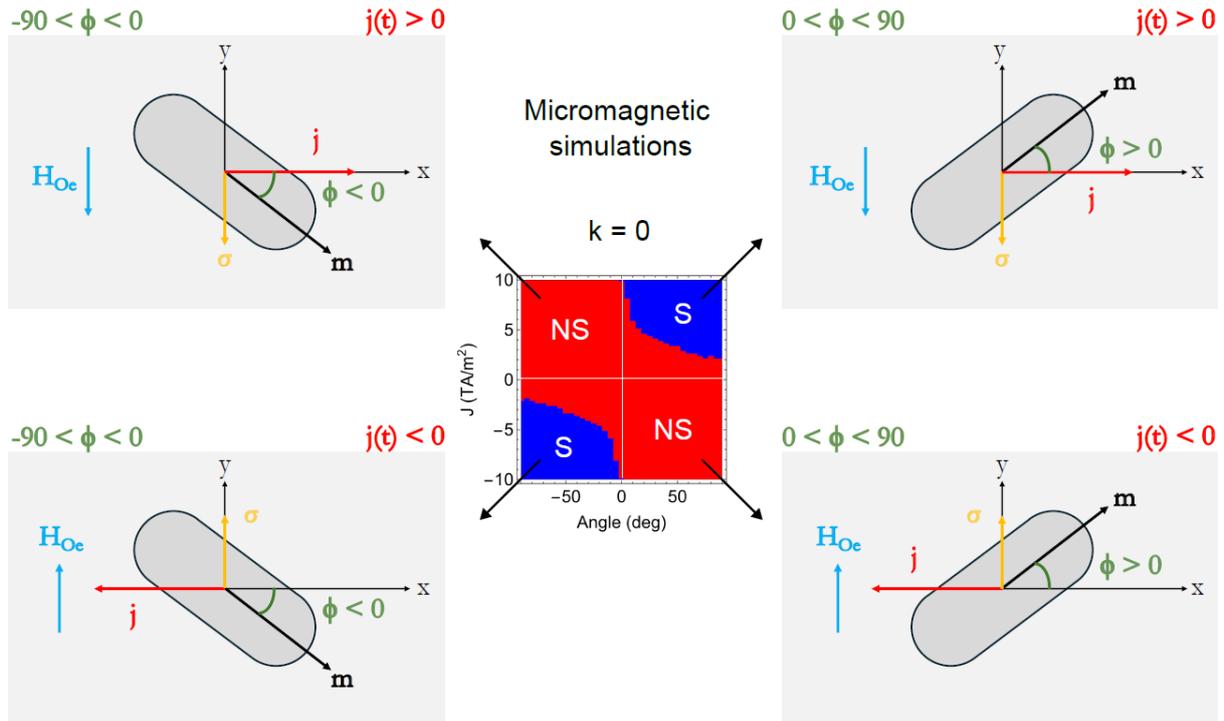

**Fig. S6.1 | Micromagnetic simulations across the switching map of magnetization angle ($\phi$) versus current density ($J$).** The central panel presents the switching behavior of a single nanomagnet as a function of the angle $\phi$ between the magnetization and the current direction, and the applied current density ($J$). The four different schematics around the central panel depict the different orientations of the nanomagnet and the current polarity. The simulations are performed for the case of no field-like torque, i.e. $k = 0$. In the switching map, blue regions, labelled "S" for "switching", correspond to successful magnetization reversal. Red regions indicate that the magnetization remains unchanged, i.e. not switched ("NS").

To complement our experimental data, we analyzed the dependence of $J_{switch}$ on $\phi$ using micromagnetic simulations (Fig. S6.2). We first compare the overall switching trend from Fig. 1(e) with the simulation results in Fig. S6.2 (a), which include the Oersted field and varying ratios of field-like and damping-like torque components. Generally, the $J_{switch}$ ($\phi$) dependence and the angle $\phi$ at which the $J_{switch}$ attains a minimum can be modified by $k$. When $k = -0.5$, switching is absent for $\phi < 40°$, which is not seen in the experimental data. The less pronounced minimum for $k=+0.5$ suggests that the observed switching behavior arises from a combination of both torque components rather than a single dominant one.

In Fig. S6.2 (b), the micromagnetic simulations are repeated, but this time without any Oersted field. In this case, the $k = -0.5$ data lose a clear switching trend, while for $k = 0$, switching becomes irregular above 45°. For $k = 0.5$, the switching minimum shifts to larger angles than those observed experimentally. These results indicate that in our system, all contributions—the field-like torque, damping-like torque, and the Oersted field—determine the switching behavior observed in Fig. 1(e).

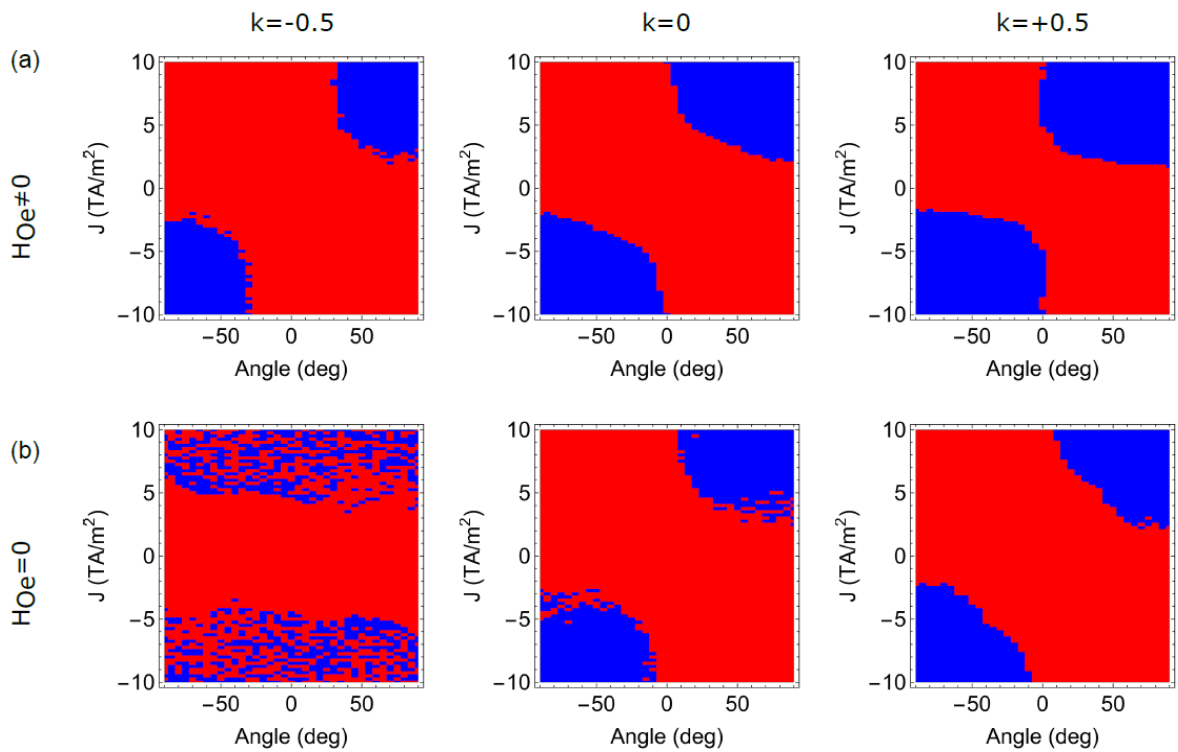

**Fig. S6.2 | SOT switching of a single nanomagnet determined from micromagnetic simulations.** (a) Switching maps based on the results from micromagnetic simulations including the Oersted field for different. The direction of the Oersted field depends on the direction of the applied current, as illustrated in Fig. S6.1. (b) The same micromagnetic simulations without including the Oersted field.

[1]   A. Vansteenkiste, J. Leliaert, M. Dvornik, M. Helsen, F. Garcia-Sanchez, and B. Van Waeyenberge, The design and verification of MuMax3, AIP Advances **4**, 107133 (2014).